\pdfoutput=1
\documentclass[11pt]{article}
\usepackage{cite}
\usepackage{graphicx}
\usepackage{latexsym}   
\usepackage{mathrsfs}
\usepackage[scr=rsfs,cal=boondox]{mathalfa}
\usepackage[overload]{textcase}

\font\grande=cmr9.5 scaled \magstep4
\font\medio=cmr9.5 scaled \magstep2
\outer\def\beginsection#1\par{\medbreak\bigskip
      \message{#1}\leftline{\bf#1}\nobreak\medskip
\vskip-\parskip
      \noindent}

\begin{document}
\bibliographystyle{unsrt}

\titlepage

\vspace{1cm}
\begin{center}
{\grande Relic gravitons and non-stationary processes}\\
\vspace{0.5cm}
\vspace{1.5 cm}
Massimo Giovannini \footnote{e-mail address: massimo.giovannini@cern.ch}\\
\vspace{0.5cm}
{{\sl Department of Physics, CERN, 1211 Geneva 23, Switzerland }}\\
\vspace{0.5cm}
{{\sl INFN, Section of Milan-Bicocca, 20126 Milan, Italy}}\\
\vspace*{1cm}
\end{center}
\vskip 0.3cm
\centerline{\medio  Abstract}
\vskip 0.5cm
Stationary processes do not accurately describe the diffuse backgrounds of relic gravitons whose correlations are homogeneous in space (i.e. only dependent upon the distance between the two spatial locations) but not in time. The symmetries of the autocorrelations ultimately reflect the quantum mechanical origin of the diffuse backgrounds and lead to non-stationary observables at late time. In  particular, large oscillations are believed to arise in the spectral energy density that is customarily (but approximately) related to the tensor power spectrum. When the full expression of the spectral energy density is employed the amplitudes of oscillation are instead suppressed in the large-scale limit and the non-stationary features of the late-time signal practically disappear. For similar reasons the relations between the spectral energy density and the spectral amplitude are ambiguous in the presence of non-stationary features. While it is debatable if  the non-stationary features are (or will be) directly detectable, we argue that the spectral amplitude following from the Wiener-Khintchine theorem is generally inappropriate for a consistent description of the relic signal. Nevertheless the strong oscillatory behaviour of the late-time observables is naturally smeared out provided the spectral energy density is selected as pivotal variable.
\noindent
\vspace{5mm}
\vfill

\newpage
\renewcommand{\theequation}{1.\arabic{equation}}
\setcounter{equation}{0}
\section{Introduction}
\label{sec1}
The relic gravitational waves produced by the early variation of the space-time curvature \cite{GG1,GG2,GG3,GG4} lead to a late-time background of diffuse radiation. In the simplest situation the relic gravitons are produced in pairs of opposite three-momenta from the inflationary vacuum and this is why they appear as a collection of standing (random) waves which are the tensor analog of the so-called Sakharov oscillations \cite{GG5}; this phenomenon has been also independently discussed in the classic paper of Peebles and Yu \cite{GG6} (see also \cite{GG7}). The question analyzed in this paper involves the possibility of describing the relic gravitons in terms of a stationary and homogeneous stochastic process. As we shall see the answer to this question depends, to some extent, on the pivotal variable that is selected for the physical description of the relic graviton background.

A first general observation relevant in this context is that the late-time properties of the signal not only rest on the features of the inflationary vacuum but also on the post-inflationary evolution. It is well established that in the concordance paradigm the spectral energy density at late times is quasi-flat \cite{SS1,SS1a,SS1b} and it is maximized\footnote{In the concordance scenario the spectral energy density 
    scales as $\nu^{-2}$ between few aHz and $100$ aHz. This means that it is larger when the frequency is comparatively smaller.} in the aHz region\footnote{As usual we employ the prefixes of the international system of units so that, for instance, $1\, \mathrm{aHz} = 10^{-18}\,\, \mathrm{Hz}$, $1 \, \mathrm{nHz} = 10^{-9} \,\, \mathrm{Hz}$ and so on and so forth.} \cite{SS2,SS3} where the current Cosmic Microwave Background (CMB) observations are now setting stringent limits on the contribution of the relic gravitons to the temperature and polarization anisotropies \cite{TS1,TS2,TS3}.  The low-frequency constraints can be viewed as direct bounds on the tensor to scalar ratio $r_{T}$ and seem to suggest that at higher frequencies (i.e. in the audio band and beyond) the spectral energy density in critical units should be ${\mathcal O}(10^{-17})$ or even smaller.  The minuteness of the spectral energy density is however based on the presumption that radiation dominates (almost) right after the end of inflation and it is otherwise invalid \cite{SS6}. The post-inflationary evolution prior to BBN nucleosynthesis is not probed by any direct observation and it can deviate from the radiation dominated evolution; if this is the case, it has been argued long ago that the high-frequency spectrum of the relic gravitons can be much larger \cite{SS6} (see also Ref.  \cite{SS7} for a recent review). Even though the detectability of the signal is essential, for the present ends what matters are mainly the symmetries of the correlation functions.
 
The second general remark is that, at the moment, the direct observations that are potentially relevant for the relic graviton backgrounds involve the pulsar timing arrays (PTA) in the nHz range \cite{PPTA2,NANO2,EPTA2,CPTA} and the ground-based interferometers \cite{LIGO1,LIGO2} operating in the audio band. It is actually well established since the late 1970s that the millisecond pulsars can be employed as effective detectors of random gravitational waves for a typical frequency domain that corresponds to the inverse of the observation time during which the pulsar timing has been monitored (see e.g. \cite{PP1a,PP1b,PP1c}). The correlation signature of an isotropic and random gravitational wave background should follow the so-called Hellings-Downs\footnote{In case the gravitational waves {\em are not} characterized by stochastically distributed Fourier amplitudes the corresponding signal does not necessarily follow the Hellings-Downs correlation.  The previous data releases of the PTAs did not report specific evidence on the Hellings-Downs correlation \cite{NANO1,PPTA1,EPTA1,IPTA1}. 
The last data releases seem to suggest more compelling 
evidences \cite{PPTA2,NANO2,EPTA2,CPTA} even if two competing experiments \cite{PPTA2,NANO2} make slightly different statements on the Hellings-Downs correlation.} curve \cite{PP1c}.  A particularly interesting aspect, for the present purposes, involves the time-dependence of the signal suggested in \cite{PPTA2}. At much higher frequencies the wide-band detectors are now setting bounds on the diffuse backgrounds of gravitational radiation between few Hz and $10$ kHz. Since the late 1990s these bounds have been greatly improving and are now broadly compatible both with the PTA observations and with the big-bang nucleosynthesis constraints \cite{bbn1,bbn2,bbn3}.
In the future it would be interesting to have detectors of relic gravitons operating directly in the MHz or GHz regions \cite{cav1b,cav2,cav2a,cav3,cav4,cav5a,cav6,cav6a,cav7,cav8} where direct bounds on relic gravitons 
are essential for determining the early expansion history
of the Universe \cite{HFD}.

All in all, taking into account the considerations of the two previous paragraphs, the current searches of diffuse backgrounds of gravitational radiation assume that signal is in fact described by a stationary stochastic process characterized by a time-independent spectral amplitude \cite{STOC1,STOC2}. In this approach the correlation functions of the signal at different instants depend on the difference between the two times at which the random fields are evaluated so that the spectral amplitude is simply given by the Fourier transform of the autocorrelation function \cite{WK1,WK2}.  As far as the spatial dependence is concerned the stochastic process is usually assumed to be homogeneous; by this we mean, according to the usual terminology, that the ensemble averages of the random fields evaluated at two different points depend on the distance between the two points. If the relic gravitons are produced from the quantum fluctuations of the gravitational field the homogeneity of the process is natural but not its stationarity. 

The production of relic gravitons stipulates that the initial quantum state has evolved into a correlated multiparticle state \cite{relic}. As a consequence, in the  Heisenberg description the field operators exhibit a characteristic pattern of non-stationary and standing oscillations which are in fact the tensor analog of the well known Sakharov oscillations \cite{GG5}. When estimating the signal to noise ratio associated with the diffuse backgrounds of gravitational radiation the properties of the relic signal are assumed to be similar to the ones of the intrinsic noises of the detectors namely Gaussian, uncorrelated, stationary and statistically independent on the possible presence of other diffuse backgrounds \cite{CORR1,CORR2,CORR3,CORR4, CORR5, CORR6,CORR7,CORR7a}.  In the first part of the present analysis, following some earlier observations \cite{CORR5,CORR6} we shall first clarify that the relic graviton backgrounds are per se not equivalent to a stationary stochastic process since their autocorrelation function does not only depend on the time difference. In the second part of this investigation 
we shall present a number of specific examples. Finally in the third portion of the paper we shall 
analyze the variables where the non-stationary features are less pronounced. Contrary to a naive intuition we shall argue that the spectral amplitude is not appropriate for the description of a non-stationary signal while the spectral energy density is far more convenient. However the most heuristic approaches used for the estimate of the spectral energy density simply assume a well defined relation between the tensor power spectrum and the energy density without appreciating that such a relation is indeed approximate. As a consequence the spectral energy density exhibits late-time oscillations that are instead spurious. The best strategy for a direct evaluation of the spectral energy density is instead to impose the small-scale limit only after assessing all the relevant power spectra in their exact form. When this is properly done, the oscillations of the spectral energy density are suppressed at late times and $\Omega_{gw}(\nu,\tau)$ turns out to be quasi-stationary; by this we mean that the strong oscillations do not arise to leading order but only in the corrections that are irrelevant for all the wavelengths shorter than the Hubble radius.

The layout of the paper is, in short, the following. In section \ref{sec2} the stochastic  
processes are swiftly introduced with a particular focus on the tensor random fields. In section \ref{sec3} we discuss  the case of the relic gravitons and demonstrate that they cannot be reduced to the case of a stationary random process discussed in section \ref{sec2}. For this purpose  the correlation functions are defined in terms of the appropriate quantum field operators. In section \ref{sec4} the late-time autocorrelation functions are evaluated in two physically significant cases: namely the situations where the universe is dominated by radiation and dusty matter.  In section \ref{sec5} we analyze the physical features of the spectral energy density 
and show that the order $1$ oscillations are not a consequence 
of the non-stationary nature of the diffuse background but rather of the approximation scheme.
We then compute exactly the spectral energy density in all the physical situations 
exemplified in section \ref{sec4} and conclude that the spectral energy density 
only contains oscillating corrections that are suppressed in the large-scale limit.
At the end of section \ref{sec5}  the spectral amplitude and the chirp amplitude are more specifically 
considered. The concluding remarks are collected in section \ref{sec6}.
We found useful to relegate to the appendix a number of relevant technical discussions that could
otherwise look as unwanted digressions in the bulk of the paper. In particular we discussed in appendix \ref{APPA} the case of scalar random fields; appendix \ref{APPB} is instead devoted to the 
explicit forms of the transition matrices that are employed in sections \ref{sec4} and \ref{sec5}. 

\renewcommand{\theequation}{2.\arabic{equation}}
\setcounter{equation}{0}
\section{Stationary and non-stationary stochastic processes}
\label{sec2}
The autocorrelation function of a {\em stationary} random process only depends upon the difference between the times at which the random fields of the ensemble 
average are evaluated \cite{STOC1,STOC2}. The autocorrelation function must then be invariant under a common shift of the time coordinate and, for this reason, its Fourier transform is associated with a well defined spectral amplitude \cite{WK1,WK2}. A stochastic process is instead {\em homogeneous} when the correlations of the relevant random fields evaluated at different points only depend upon the distance between the two spatial coordinates. Both stationarity and homogeneity play an important role when 
analyzing the correlation between gravitational wave detectors of arbitrary geometry \cite{CORR1,CORR2,CORR3,CORR4}. In particular the intrinsic noises of the instruments are customarily assumed to be  stationary, Gaussian, uncorrelated, much larger in amplitude than the gravitational strain, and statistically independent on the strain itself.  The stationarity and the homogeneity are also conjectured 
 for the signals associated with the diffuse background of gravitational radiation \cite{CORR7}. In what follows, after presenting the stationarity in the case of an ensemble of random functions, we consider the case of tensor random fields. To avoid digressions  the results of the scalar case (employed in some of the derivations of this section) have been collected in the appendix \ref{APPA}. 

\subsection{Random functions and stationary processes}
Let us consider, for the sake of simplicity, an ensemble of real random functions $q(\tau)$ 
where $\tau$ denotes throughout the (conformal) time coordinate of the problem
even if in this section the accurate identification of $\tau$ will not be essential\footnote{In this section the curvature 
of the space-time does not play any role. However, to avoid confusions, $\tau$ denotes throughout the conformal time coordinate. The background geometry will then be assumed to be conformally flat and characterized by a scale 
factor $a(\tau)$ so that the relations between the cosmic and the conformal time coordinates is given by $a(\tau) d\tau = d t$.}.  With these specifications, the autocorrelation function can be defined in the context of the generalized harmonic analysis and its existence is associated with the finiteness of the integral \cite{STOC1}
\begin{equation}
\Gamma_{q}( \Delta\tau ) = \lim_{T \to \infty}\,\, \frac{1}{2 \, T}\, \int_{-T}^{T} \, q(\tau) q(\tau + \Delta\tau) \, d \tau,
\label{ST1}
\end{equation}
defined in the Lebesgue sense. The expression of Eq. (\ref{ST1}) applies in the case 
of a single function $q(\tau)$ and does not refer to any statistical concept. When dealing 
with a  stationary and ergodic ensemble of random functions, the autocorrelation 
of Eq. (\ref{ST1}) can be replaced by 
\begin{equation}
\Gamma_{q}(|\tau_{1}- \tau_{2}|) = \langle \, q(\tau_{1}) \, q(\tau_{2})\, \rangle,
\label{ST2}
\end{equation}
where $\langle \,.\,.\,.\rangle$ now denotes an ensemble average whose result 
coincides, by definition, with Eq. (\ref{ST1}) because of the hypotheses 
of ergodicity and stationarity. As we shall see later on in this section Eqs. (\ref{ST1})--(\ref{ST2}) are easily 
generalized to the case of a stochastic quantum field. The Fourier transform of the autocorrelation function 
is ($S_{q}(\nu)$ in what follows) is, by definition,  the spectral amplitude of the process: 
\begin{equation}
q(\tau) = \int_{-\infty}^{+\infty} e^{ 2 \, i\, \pi\, \nu \, \tau} q(\nu) \, d\nu, \qquad \langle q(\nu)\, q(\nu^{\prime}) \rangle = \delta(\nu + \nu^{\prime}) \, S_{q}(\nu),
\label{ST2c}
\end{equation}
where $\nu$ is the frequency\footnote{We shall use throughout the natural units $\hbar= c= 1$;
 this means, in particular, that $\omega = k= 2 \pi \,\nu$ where $\nu$ is the frequency and $\omega$ the angular frequency. Occasionally in the literature involving the diffuse backgrounds of gravitational radiation the frequency $\nu$ is also denoted by $f$ (see e.g. \cite{CORR7}) but we shall not use this notation that would be ambiguous in the present context.}. The autocorrelation function and the spectral amplitudes are then related as 
\begin{eqnarray}
\Gamma_{q}(\tau_{1} - \tau_{2}) = \frac{1}{2\pi} \int_{-\infty}^{\infty} e^{i \, \omega (\tau_{1} - \tau_{2})} S_{q}(\omega) d \omega 
= \int_{-\infty}^{\infty} e^{2 i \,\pi \, \nu (\tau_{1} - \tau_{2})} \, S_{q}(\nu) \, d \nu.
\label{ST2d}
\end{eqnarray}
According to Eq. (\ref{ST2d}) the spectral amplitude and the autocorrelation function of the process 
form a Fourier transform pair; this statement is often referred to as Wiener-Khintchine (see e.g. \cite{STOC2}) 
theorem and was originally developed in the framework of the so-called generalized harmonic analysis  that 
establishes a rigorous connection between Eqs. (\ref{ST1}) and (\ref{ST2}) \cite{WK1,WK2}. The possibility of defining a spectral amplitude relies on the stationary nature of the underlying random process. The nomenclature 
employed hereunder is the one established in Eq. (\ref{ST2d}) and we shall call $S_{q}(\nu)$ {\em spectral amplitude}. There are however other terminologies: some authors call $S_{q}(\nu)$ {\em spectral density} or even power spectrum. For the sake of accuracy we stress that, in the present context, the power spectrum can be related to $S_{q}(\nu)$ in the case of a stationary process but it is, generally speaking, a different quantity. In particular the power spectrum defined hereunder is dimensionless whereas the spectral amplitude $S_{q}(\nu)$ has dimensions 
of an inverse frequency (or of a time).

 \subsection{Tensor random fields}
 \subsubsection{Stationary processes}  
A solenoidal (and traceless) tensor random field can be treated with the same strategy already 
introduced in the scalar case. In particular the tensor amplitude can be transformed as
\begin{equation}
h_{i\, j}(\vec{x}, \tau) = \int_{-\infty}^{\infty} d \nu \int d\,\widehat{k}  \, e^{ 2\,i\,\pi \, \nu\,( \tau - \widehat{k}\cdot\vec{x})} \,h_{i\, j}(\nu, \widehat{k}), 
\label{STR3aa}
\end{equation}
where $h^{*}_{i\, j}(\nu, \widehat{k}) = \,h_{i\, j}(-\nu, \widehat{k})$ 
and $\nu$ denotes, as usual, the comoving frequency.  The tensor amplitude $h_{i\, j}(\nu, \widehat{k})$ can be expanded in the basis of the linear  polarizations. As usual we introduce three orthogonal unit vectors 
$\hat{m}$, $\hat{n}$ and $\hat{k}$ so that the two tensor polarizations as $e_{i\,j}^{\oplus}= (\hat{m}_{i} \, \hat{m}_{j} - 
\hat{n}_{i}\, \hat{n}_{j})$ and $e_{i\,j}^{\otimes}= (\hat{m}_{i} \, \hat{n}_{j} + 
\hat{n}_{i}\, \hat{m}_{j})$. With these standard notations we can write
\begin{equation}
h_{i\, j}(\nu, \widehat{k}) = \sum_{\lambda= \oplus, \, \otimes} \, e_{i\, j}^{(\lambda)}(\widehat{k}) \, h_{\lambda}(\nu, \widehat{k}). 
\label{STR3ab}
\end{equation}
In analogy with the scalar case $S_{h}(\nu)$ is introduced from the expectation value of the tensor amplitudes
expressed as a function of $\nu$ and $\widehat{k}$:
\begin{equation}
\langle h_{\lambda} (\nu, \, \widehat{k}) \, h_{\lambda^{\prime}} (\nu^{\prime}, \, \widehat{k}^{\prime}) \rangle = 
{\mathcal C}_{h}\,\, S_{h}(\nu) \,\, \delta(\nu + \nu^{\prime})\, \, \delta^{(2)}(\widehat{k} - \widehat{k}^{\,\prime}) \,\, \delta_{\lambda\,\lambda^{\prime}},
\label{STR3ac}
\end{equation}
where ${\mathcal C}_{h}$ is an overall constant that plays the same role of ${\mathcal C}_{\phi}$ introduced in the scalar case (see appendix \ref{APPA} and, in particular, Eqs. (\ref{ST2e0}), (\ref{ST2h}) and (\ref{ST2s})). From  Eq. (\ref{STR3ac}) the autocorrelation function $\Gamma_{h}(\tau_{1} - \tau_{2})$ is introduced from:
\begin{equation}
\langle h_{\lambda}(\widehat{k}, \tau_{1}) \, h_{\lambda^{\prime}}(\widehat{p}, \tau_{2}) \rangle = {\mathcal C}_{h} \, \delta_{\lambda\lambda^{\prime}}\,
\delta^{(2)}(\widehat{k} - \widehat{p}) \, \Gamma_{h}(\tau_{1} - \tau_{2}).
\label{STR3ad}
\end{equation}
Ultimately the connection between the spectral amplitude $S_{h}(\nu)$ and the autocorrelation function $\Gamma_{h}(\tau_{1} - \tau_{2})$ is given by:
\begin{equation}
S_{h}(\nu) = \int_{-\infty}^{\infty} e^{ 2 \, i\, \nu\, z} \, \Gamma_{h}(z) \, d z,
\label{STR3ae}
\end{equation}
in full analogy with the scalar case of Eq. (\ref{ST2h}). 
We can finally evaluate the expectation value of two tensor amplitudes with different indices; from Eqs. (\ref{STR3ab})--(\ref{STR3ac}) we obtain:
\begin{eqnarray}
\langle h_{i\, j}(\nu, \widehat{k}) \,\,h_{\ell\, m}(\nu^{\prime}, \widehat{k}^{\prime}) \rangle = 4 {\mathcal C}_{h} {\mathcal S}_{i\,j\,m\,n}(\hat{k}) 
S_{h}(\nu) \, \delta^{(2)}(\widehat{k} - \widehat{k}^{\prime}) \, \delta(\nu + \nu^{\prime}).
\label{STR3af}
\end{eqnarray}
In Eq. (\ref{STR3af}) the sum over the polarizations has been 
expressed in terms of ${\mathcal S}_{i\, j\, m\, n}$  
\begin{equation}
{\mathcal S}_{i\, j\, m\, n} = \bigl[p_{i \,m}(\hat{k}) p_{j\, n}(\hat{k}) + p_{i \,n}(\hat{k}) p_{j\, m}(\hat{k}) - 
p_{i \,j}(\hat{k}) p_{m\, n}(\hat{k}) \bigr]/4.
\label{STR3ag}
\end{equation}
where, as usual,  $p_{i\, j}(\hat{k}) = ( \delta_{i\, j} - \widehat{k}_{i} \, \widehat{k}_{j})$ denotes the transverse projector.

\subsubsection{Homogeneous processes}
Using Eq. (\ref{STR3af}) and the observation that  ${\mathcal S}_{i\,j\,i\,j} =1$, the expectation value of the tensor amplitudes at equal time becomes
\begin{equation}
\langle h_{i\,j}(\vec{x}, \tau)\, h^{i \, j}(\vec{x}+ \vec{r},\tau) \rangle = 32 \,\pi\, {\mathcal C}_{h} 
\int_{0}^{\infty} d \nu \, S_{h}(|\nu|) \, j_{0}(2 \pi \,|\nu|\,r).
\label{STR3ai}
\end{equation}
The same strategy illustrated in the scalar case suggests that the Fourier amplitudes 
of the tensor modes can be related to the tensor power spectrum conventionally denoted by $P_{T}(k,\tau)$.
For this purpose we write
\begin{equation}
h_{i\, j}(\vec{x}, \tau) = \frac{1}{(2\pi)^{3/2}} \int d^{3} k e^{- i \, \vec{k}\cdot\vec{x}} \, \overline{h}_{i\,j}(\vec{k}, \tau), \qquad \qquad 
\overline{h}_{i\,j}(\vec{k},\tau) = \overline{h}_{i\,j}^{\ast}(-\vec{k}, \tau).
\label{STR3am}
\end{equation}
In analogy with the scalar case, for a homogeneous stochastic process the Fourier amplitudes obey
\begin{equation}
\langle \overline{h}_{i\,j}(\vec{k}, \tau) \,\, \overline{h}_{m\,n}(\vec{p}, \tau) \rangle = \frac{2 \pi^2}{k^3} \delta^{(3)}(\vec{k}+ \vec{p}) P_{T}(k,\tau) {\mathcal S}_{i\, j\,m\,n}(\hat{k}).
\label{STR3an}
\end{equation}
It then follows that the expectation value of the quadratic combination 
$h_{i\,j}(\vec{x}, \tau)\, h^{i \, j}(\vec{x}+\vec{r},\tau)$ can be written in two different ways depending the expansion we use; but since 
both expansions refer to the same tensor random field they must ultimately coincide so that 
\begin{equation}
 \langle \,h_{i\,j}(\vec{x}, \tau)\, h^{i \, j}(\vec{x} +\vec{r},\tau) \,\rangle = 32 \, \pi\, {\mathcal C}_{h} 
\int_{0}^{\infty} d \nu \, S_{h}(|\nu|)  \, j_{0}(2 \pi \,|\nu|\,r) = \int_{0}^{\infty} \frac{d\, k}{k} P_{T}(k,\tau) \, j_{0}(k\,r).
\label{examplecon2}
\end{equation}
Again, provided the tensor power spectrum is truly stationary we can then relate $P_{T}(k)$ to the 
spectral amplitude with the result that, as expected,
\begin{equation}
\nu \, S_{h}(|\nu|) = P_{T}(\nu), \qquad 32 \,\pi\, {\mathcal C}_{h} = 1,
\label{CONN1}
\end{equation}
where the difference between the condition appearing in Eq. (\ref{ST2s}) comes from the 
sum over the polarizations. As in the scalar case, 
the connection between the spectral amplitude and the tensor power spectrum obtained in Eq. (\ref{CONN1})
is only rigorous when the stochastic process is {\em both} stationary and homogeneous.

\renewcommand{\theequation}{3.\arabic{equation}}
\setcounter{equation}{0}
\section{The relic gravitons and their quantum correlations}
\label{sec3}
\subsection{Stochastic processes and quantum expectation values}
In a conformally flat background geometry chracterized by a scale factor $a(\tau)$ (where $\tau$ now denotes the conformal time coordinate) the tensor modes of the geometry may be amplified from their quantum mechanical fluctuations \cite{GG1,GG2,GG3,GG4}.  For the sake of illustration we shall be considering an inflationary stage possibly followed by the standard post-inflationary evolution  \cite{SS1,SS1a,SS1b,SS2,SS3}. When the scalar and tensor modes 
of the geometry are amplified from there quantum fluctuations (see, for instance, \cite{SS7}) the random fields introduced 
in section \ref{sec2} must be replaced by the appropriate field operators $\widehat{h}_{i\,j}(\vec{x},\tau)$ that 
are solenoidal (i.e. $\partial_{i} \, \widehat{h}^{\,\,i}_{j} =0$) and traceless (i.e. $\widehat{h}_{i}^{\,\,i} =0$). The expectation values of these field operators in Fourier space define the tensor power spectrum
\begin{equation}
\langle \, \widehat{h}_{i\, j}(\vec{k},\tau) \, \, \widehat{h}_{m\, n}(\vec{p},\tau) \rangle = \frac{2 \pi^2}{k^3} \, {\mathcal S}_{i\, j \, m\,n}(\hat{k}) P_{T}(k, \tau) \, \delta^{(3)}(\vec{k} + \vec{p}),
\label{ST4}
\end{equation}
which is in fact the analog of Eq. (\ref{STR3an}) with the difference that
its specific form now depends on the mode functions of the quantum field. The expectation value  of the tensor amplitude must be complemented by the one of the corresponding time derivatives whose evolution cannot be neglected:
\begin{equation}
\langle \, \partial_{\tau}\, \widehat{h}_{i\, j}(\vec{k},\tau) \, \, \partial_{\tau} \widehat{h}_{m\, n}(\vec{p},\tau) \rangle = \frac{2 \pi^2}{k^3} \, {\mathcal S}_{i\, j \, m\,n}(\hat{k}) Q_{T}(k, \tau) \, \delta^{(3)}(\vec{k} + \vec{p}),
\label{ST6}
\end{equation}
where $Q_{T}(k,\tau)$ is the corresponding power spectrum. The explicit form of the field operators in the Heisenberg representation is: 
\begin{equation}
\widehat{h}_{i\, j}( \vec{x}, \tau) = \frac{\sqrt{2} \, \ell_{P}}{(2 \pi)^{3/2}} \sum_{\alpha = \oplus, \otimes} \int d^{3} k \,\, e^{(\alpha)}_{i\,j}(\hat{k}) \,\,
\biggl[ F_{k,\alpha}(\tau) \, \widehat{b}_{\vec{k}, \, \alpha} e^{- i \vec{k}\cdot \vec{x}} + \mathrm{H.\,c.}\biggr],
\label{ST7}
\end{equation}
where $\ell_{P} = 8 \pi G$ is the Planck length. In Eq. (\ref{ST7}) the second term inside the square bracket denotes the Hermitian conjugate of the preceding one and the sum runs over the two orthogonal tensor polarizations defined previously (see Eq. (\ref{STR3aa}) and discussion thereafter).  Sticking to the mode expansion of Eq. (\ref{ST7}) the corresponding canonical momenta are given by\footnote{Both $\widehat{h}_{i\,j}(\vec{x}, \tau)$ and $\widehat{\pi}_{i\,j}(\vec{x}, \tau)$ are Hermitian, i.e. 
$\widehat{h}_{i\,j}^{\, \dagger}(\vec{x}, \tau) = \widehat{h}_{i\,j}(\vec{x}, \tau)$ and $\widehat{\pi}_{i\,j}^{\, \dagger}(\vec{x}, \tau) = \widehat{\pi}_{i\,j}(\vec{x}, \tau)$.}: 
\begin{equation}
\widehat{\pi}_{i\, j}( \vec{x}, \tau) = \frac{a^2(\tau)}{4 \sqrt{2} \, \ell_{P}\, (2 \pi)^{3/2}} \sum_{\beta = \oplus, \otimes} \int d^{3} k \,\, e^{(\beta)}_{i\,j}(\hat{k}) \,\,
\biggl[ G_{k,\beta}(\tau) \, \widehat{b}_{\vec{k}, \, \beta} e^{- i \vec{k}\cdot \vec{x}} + \mathrm{H.\,c.}\biggr],
\label{ST7a}
\end{equation}
where $a(\tau)$ is the scale factor; the mode functions for the momenta are $G_{k}= F_{k}^{\prime}$ where the prime will denote, from now on, the derivation with respect to the conformal time coordinate $\tau$.  
The Fourier transform of the Hermitian field operators of Eqs. (\ref{ST7})--(\ref{ST7a}) is then given by 
\begin{eqnarray}
\widehat{h}_{i\,j}(\vec{q},\tau) &=& \sqrt{2}\, \ell_{P} \, \sum_{\alpha} \biggl[ e^{(\alpha)}_{i\,j}(\hat{q})\,\widehat{b}_{\vec{q},\, \alpha} \, F_{q,\alpha}(\tau) +  e^{(\alpha)}_{i\,j}(-\hat{q})\widehat{b}_{-\vec{q}, \alpha}^{\dagger} F_{q,\alpha}^{\ast}(\tau)\biggr],
\label{ST7b}\\
\widehat{\pi}_{m\,n}(\vec{p},\tau) &=& \frac{a^2}{4\,\sqrt{2}\, \ell_{P}}\, \sum_{\beta} \biggl[ e^{(\beta)}_{m\,n}(\hat{p})\,\widehat{b}_{\vec{p},\, \beta} \, G_{p,\beta}(\tau) +  e^{(\beta)}_{m\,n}(-\hat{p})\widehat{b}_{-\vec{p}, \beta}^{\dagger} G_{p,\beta}^{\ast}(\tau)\biggr].
\label{ST7c}
\end{eqnarray}
If we then impose canonical commutation relations between the field operators of Eqs. (\ref{ST7b})--(\ref{ST7c}) 
\begin{equation}
\biggl[ \widehat{h}_{i\, j}(\vec{q},\tau), \, \widehat{\pi}_{m\, n}(\vec{p}, \tau) \biggr] = i\,\, {\mathcal S}_{i\,j\, m\, n}(\widehat{q}) \,\, \delta^{(3)}(\vec{q} + \vec{p}),
\label{ST7d}
\end{equation}
must necessarily be normalized as:
\begin{equation}
F_{q}(\tau) \, G_{q}^{\ast}(\tau) - F_{q}^{\ast}(\tau) G_{q}(\tau) = i/a^2(\tau).
\label{ST7e}
\end{equation}
The condition expressed by Eq. (\ref{ST7e}) is essential to obtain the correct final expressions of the mode functions and it is verified throughout all the stages of the dynamical evolution.

\subsection{Homogeneous processes and generalized autocorrelation functions}
If the explicit expressions of Eqs. (\ref{ST7b})--(\ref{ST7c}) are inserted into Eqs. (\ref{ST4})--(\ref{ST6}) the two power spectra $P_{T}(k,\tau)$ and $Q_{T}(k,\tau)$ are given by:
\begin{equation}
P_{T}(k,\tau) = \frac{4 \ell_{P}^2}{\pi^2}\, k^3\, \bigl| F_{k}(\tau) \bigr|^2, \qquad Q_{T}(k,\tau) = \frac{4 \ell_{P}^2}{\pi^2}\, k^3\, \bigl| G_{k}(\tau) \bigr|^2.
\label{ST8}
\end{equation}
The two-point functions 
associated with $P_{T}(k,\tau)$ and $Q_{T}(k,\tau)$ are in fact associated with 
homogeneous two-point functions of the type of Eq. (\ref{examplecon2})
\begin{eqnarray}
&& \langle \widehat{h}_{i\, j}(\vec{x}, \tau) \,\, \widehat{h}^{i\, j}(\vec{x} + \vec{r}, \tau) \rangle = \int_{0}^{\infty} P_{T}(k,\tau) \, j_{0}(k \, r) \, \frac{d k}{k},
\nonumber\\
&& \langle \partial_{\tau}\widehat{h}_{i\, j}(\vec{x}, \tau) \,\, \partial_{\tau}\widehat{h}^{i\, j}(\vec{x} + \vec{r}, \tau) \rangle = \int_{0}^{\infty} Q_{T}(k,\tau) \, j_{0}(k \, r) \, \frac{d k}{k}.
\label{ST8a}
\end{eqnarray}
To analyze the stationarity of the process we therefore need to introduce 
the autocorrelation functions that we define as:
\begin{eqnarray}
\Gamma_{i\,j\,m\,n}(\vec{k}, \vec{p}, \tau_{1}, \tau_{2}) &=& \frac{1}{2}\biggl[\langle \widehat{h}_{i\,j}(\vec{k}, \tau_{1}) \, \widehat{h}_{m\,n}(\vec{p}, \tau_{2}) \rangle  +  \langle \widehat{h}_{i\,j}(\vec{p}, \tau_{2}) \, \widehat{h}_{m\,n}(\vec{k}, \tau_{1}) \rangle\biggr],
\label{ST9}\\
\overline{\Gamma}_{i\,j\,m\,n}(\vec{k}, \vec{p}, \tau_{1}, \tau_{2}) &=& \frac{1}{2}\biggl[\langle \partial_{\tau_{1}} \widehat{h}_{i\,j}(\vec{k}, \tau_{1}) \,\, \partial_{\tau_{2}} \widehat{h}_{m\,n}(\vec{p}, \tau_{2}) \rangle  +   \langle \partial_{\tau_{2}}\, \widehat{h}_{i\,j}(\vec{p}, \tau_{2}) \,\partial_{\tau_{1}} \widehat{h}_{m\,n}(\vec{k}, \tau_{1}) \rangle\biggr].
\label{ST10}
\end{eqnarray}
From Eq. (\ref{ST7}) we can deduce the explicit expressions of $\widehat{h}_{i\,j}(\vec{k}, \tau)$ and then compute directly Eqs. (\ref{ST9})--(\ref{ST10}) whose explicit form becomes 
\begin{eqnarray}
\Gamma_{i\,j\,m\,n}(\vec{k}, \vec{p}, \tau_{1}, \tau_{2}) &=& {\mathcal S}_{i\, j \, m\,n}(\hat{k})\, \delta^{(3)}(\vec{k} + \vec{p}) \, \Delta_{k}(\tau_{1},\, \tau_{2}), 
\label{ST11}\\
\overline{\Gamma}_{i\,j\,m\,n}(\vec{k}, \vec{p}, \tau_{1}, \tau_{2}) &=&  {\mathcal S}_{i\, j \, m\,n}(\hat{k}) \, \delta^{(3)}(\vec{k} + \vec{p}) \,\overline{\Delta}_{k}(\tau_{1},\, \tau_{2}) ,
\label{ST12}
\end{eqnarray}
where $\Delta_{k}( \tau_{1},\, \tau_{2})$ and $\overline{\Delta}_{k}(\tau_{1},\, \tau_{2})$ are given by:
\begin{eqnarray}
\Delta_{k}(\tau_{1},\, \tau_{2}) &=& 4 \ell_{P}^2 \biggl[ F_{k}(\tau_{1}) \,\, F_{k}^{\ast}(\tau_{2}) + F_{k}(\tau_{2}) \,\, F_{k}^{\ast}(\tau_{1}) \biggr],
\label{ST13}\\
\overline{\Delta}_{k}(\tau_{1},\, \tau_{2}) &=& 4 \ell_{P}^2 \biggl[ G_{k}(\tau_{1}) \,\, G_{k}^{\ast}(\tau_{2}) + G_{k}(\tau_{2}) \,\, G_{k}^{\ast}(\tau_{1}) \biggr].
\label{ST14}
\end{eqnarray}
The evolution of the mode functions $F_{k}(\tau)$ and $G_{k}(\tau)$ follows immediately from the action of the problem
\begin{equation}
S_{g} = \frac{1}{8 \ell_{P}^2} \int d^4 x \, \sqrt{- \overline{g}} \,\, \partial_{\mu} \, h_{i\, j}\,\, \partial_{\nu} \, h^{i\, j},
\label{ST15}
\end{equation}
where $\overline{g}_{\mu\nu}$ is the background metric and $\overline{g}$ its determinant. In the conformally flat case $\overline{g}_{\mu\nu} = a^2(\tau) \, \eta_{\mu\nu}$ and the Hamiltonian 
operator associated with the classical action (\ref{ST15}) is given by\footnote{Since the problem is inherently time-dependent, different Hamiltonians (all related by canonical transformations) can be introduced even if, ultimately, the evolution of the field operators remains unaffected. The transformed Hamiltonians might however lead to slightly different initial vacua and different canonical momenta. These aspects have been discussed in various related frameworks (see e.g. \cite{relic}) but are not central to the present discussion.}:
\begin{equation}
\widehat{H}_{g}(\tau) = \int d^{3} x \biggl[\frac{8 \,\ell_{P}^2}{a^2} \, \widehat{\pi}_{i\, j} \,\, \widehat{\pi}^{i\,j} + \frac{a^2}{8 \, \ell_{P}^2} \, \partial_{k} \widehat{h}_{i\, j} \, \partial_{k} \widehat{h}^{i\, j} \biggr].
\label{ST16}
\end{equation}
From Eq. (\ref{ST16}) the evolution equations of the field operators is then given by:
\begin{equation}
\partial_{\tau} \, \widehat{\pi}_{i\, j} = \frac{a^2}{8 \, \ell_{P}^2} \nabla^2 \widehat{h}_{i\, j}, \qquad\qquad \partial_{\tau} \, \widehat{h}_{i\, j} = \frac{8 \ell_{P}^2}{a^2} \, \widehat{\pi}^{i\, j},
\label{ST17}
\end{equation}
so that the corresponding mode functions obey: 
\begin{equation}
G_{k}^{\, \prime} + 2 {\mathcal H} \, G_{k} = - k^2 \, F_{k}, \qquad\qquad G_{k} = F_{k}^{\, \prime},
\label{ST18}
\end{equation}
where ${\mathcal H} = a^{\prime}/a$. During inflation we have that ${\mathcal H} = - 1/[ (1 - \epsilon) \, \tau]$ where $\epsilon = - \dot{H}/H^2$ is the standard slow-roll parameter.
From the Wronskian normalization condition of Eq. (\ref{ST7e}) the solution of Eq. (\ref{ST18}) 
during the inflationary stage is given by 
\begin{eqnarray}
F_{k}(\tau) &=& \frac{f_{k}}{a(\tau)}=  \frac{{\mathcal N}}{\sqrt{2 k}\,\,a(\tau)} \, \sqrt{- k\tau} \, H_{\nu}^{(1)}(- k \tau), 
\label{ST19}\\
G_{k}(\tau) &=& \frac{{\mathcal N}}{a(\tau)} \sqrt{\frac{k}{2}} \biggl[ -\frac{2 \, \nu}{\sqrt{- k \tau}} H_{\nu}^{(1)}(- k \tau) + \sqrt{- k \tau} \, H_{\nu+1}^{(1)}(- k \tau) \biggr],
\label{ST20}
\end{eqnarray}
where $H_{\nu}^{(1)}(z)$ is the Hankel function of first kind \cite{TRIC, ABR} and $\nu = (3- \epsilon)/[2( 1 -\epsilon)]$ the Bessel index; note that in Eq. (\ref{ST20}) $|{\mathcal N}| = \sqrt{\pi/2}$. It is important to remark that the initial data assigned in Eqs. (\ref{ST19})--(\ref{ST20}) are consistent 
with the Wronskian normalization 
\begin{equation}
F_{k}(\tau) G_{k}^{\ast}(\tau) - F_{k}^{\ast}(\tau) G_{k}(\tau) = i/a^2(\tau),
\label{ST20a}
\end{equation}
that must be satisfied at any stage of the dynamical evolution. Therefore the mentioned value of ${\mathcal N}$ 
implies that $F_{k}(\tau) \to e^{-i k \tau}/(\sqrt{2 k} \, a)$ and $G_{k}(\tau) = - i \sqrt{k/2} \, e^{- ik \tau}/a(\tau)$.
During an exact stage of de Sitter expansion these initial conditions are compatible with the so-called 
Bunch-Davis vacuum even though the results 
of Eqs. (\ref{ST19})--(\ref{ST20}) hold during a quasi-de Sitter stage of expansion taking place for $\tau < - \tau_{r}$,  
i.e. when the conformal time coordinate takes negative values; $\tau_{r}$ denotes, in this context 
the end of the inflationary stage and the onset of the decelerated phase. Thanks to the initial conditions of Eqs. (\ref{ST19})--(\ref{ST20}) the autocorrelation functions (\ref{ST13})--(\ref{ST14}) can then be estimated  in the different stages of the post-inflationary evolution and this is the general theme of the following section.

\renewcommand{\theequation}{4.\arabic{equation}}
\setcounter{equation}{0}
\section{The late-time values of the autocorrelation functions}
\label{sec4}
The late-time autocorrelation functions can be evaluated by enforcing at any stage of the dynamical evolution the continuity of the background and the Wronskian normalization condition of Eq. (\ref{ST7e}). These two requirements preserve the canonical form of the commutation relations and ultimately lead to standing oscillations in the mode functions. The rationale for this peculiar behaviour is related to the production of pairs of gravitons (with opposite three momenta) from the inflationary vacuum. The presence of the standing waves at late time also implies that the autocorrelation function does not only depend on the time-difference $|\tau_{1} - \tau_{2}|$, as 
as it would happen in the case of a stationary process of the kind examined in section \ref{sec1}.
We are now going to analyze the explicit form of the autocorrelation functions in two relevant examples.

 \subsection{The radiation stage and the related autocorrelation functions}
A smooth evolution of the extrinsic curvature of the background  
demands the continuity of the inflationary scale factor and of its first derivative across the transition to the radiation-dominated phase. It is practical to introduce the variable   
$u(\tau)$ accounting for the continuity of the mode functions and of the background geometry
\begin{equation}
u(\tau) = k \,[\tau + (2 - \epsilon) \tau_{r}], \qquad \tau > - \tau_{r},\qquad \epsilon = - \dot{H}/H^2,
\label{SR0}
\end{equation}
 where $\tau_{r}$ conventionally denotes the onset of the radiation-dominated stage and $\epsilon$ is the standard slow-roll parameter; by definition $u_{r} = u(-\tau_{r}) = k ( 1 -\epsilon) \tau_{r}$. It is important to stress that Eq. (\ref{SR0}) (as all the subsequent discussion) assumes a quasi-de Sitter evolution during inflation where, according to the consistency 
 conditions, $\epsilon \simeq r_{T}/16$. Even if according to observational data  $r_{T} < 0.03$ \cite{TS1,TS2,TS3}( and consequently $\epsilon < 0.001$) it is important to keep track of the solw-roll corrections since their 
 presence guarantees the continuity of the mode functions and the enforcement of the Wronskian normalization 
 condition.  In terms of $u$ and $u_{r}$ the late-time values of $f_{k}(\tau)$ and $g_{k}(\tau)$ during the radiation stage can then be expressed in terms of a $2\times2$ transition matrix 
\begin{eqnarray}
f_{k}(\tau) &=& A^{(r)}_{f\, f}(u,\, u_{r}) \overline{f}_{k}
\, +\, A^{(r)}_{f\,g}(u, \, u_{r}) \, \overline{g}_{k}/k,
\nonumber\\
g_{k}(\tau) &=& A^{(r)}_{g\,f}(u,\, u_{r}) k \, \overline{f}_{k} \, + \, A^{(r)}_{g\,g}(u,\, u_{r}) \overline{g}_{k},
\label{SR0a}
\end{eqnarray}
where, for the sake of convenience, the mode functions have been rescaled as $F_{k}(\tau) = f_{k}(\tau)/a(\tau)$ and $G_{k}(\tau) = g_{k}(\tau)/a(\tau)$. The results holding in for a radiation stage can be easily generalized to any expanding stage; we also note that the various entries of the transition matrix are all real and they are explicitly given in the appendix \ref{APPB} together with their limits. The values of $\overline{f}_{k} \equiv f_{k}(-\tau_{r})$ and $\overline{g}_{k}\equiv g_{k}(-\tau_{r})$ are fixed by the mode functions evaluated at the end of inflation (i.e. for $\tau = - \tau_{r}$) and are explicitly given by
\begin{eqnarray}
\overline{f}_{k} = \frac{{\mathcal N}}{\sqrt{2 k}} \, \sqrt{k \tau_{r}} \, H_{\nu}^{(1)}(k \tau_{r}),
\qquad
\overline{g}_{k} = {\mathcal N} \sqrt{\frac{k}{2}} \biggl[ -\frac{2 \, \nu}{\sqrt{k\tau_{r}}} H_{\nu}^{(1)}(k \tau_{r}) + \sqrt{k \tau_{r}} \, H_{\nu+1}^{(1)}(k \tau_{r}) \biggr].
\label{SR0a1}
\end{eqnarray}
As discussed in mode detail in appendix \ref{APPB} the general condition of Eq. (\ref{ST7e}) demands that 
the transition matrix is unitary i.e.
\begin{equation}
 A^{(r)}_{f\, f}(u,\, u_{r}) \, A^{(r)}_{g\,g}(u,\, u_{r}) \, - \, A^{(r)}_{f\,g}(u, \, u_{r})\, A^{(r)}_{g\,f}(u,\, u_{r})= 1.
\label{SR0a2}
\end{equation}
The condition (\ref{SR0a2}) is written in a particular case but it is obviously a general property that must be enforced 
for any continuous transition of the background geometry; see also, in this respect, the discussion after Eq. (\ref{APPA4}).
Let us now consider the modes that reentered the Hubble radius during radiation\footnote{In this case the continuity of the evolution of $a(\tau)$ and ${\mathcal H}(\tau)$ 
implies that, during radiation, $a_{r}(\tau)= [\beta(\epsilon) (\tau/\tau_{r}) + \beta(\epsilon) +1]$ 
where $\beta(\epsilon) = 1/(1-\epsilon)$. During inflation, as usual, $a \, H = - \beta(\epsilon)/\tau$ 
whereas, at the end of inflation (i.e. for $\tau\to - \tau_{r}$) $a_{r} \, H_{r}= \beta(\epsilon)/\tau_{r}$. Grossly speaking $H_{r}$ denotes the expansion rate at the end of inflation.}. 
The mode functions of Eq. (\ref{SR0a}) can be expressed in the following form
\begin{eqnarray}
f_{k}(\tau) = A^{(r)}_{f\, f}(u,\, u_{r})\,  \overline{f}_{k} \, \biggl[ 1 + \frac{A^{(r)}_{f\,g}(u, \, u_{r}) }{A^{(r)}_{f\, f}(u,\, u_{r})} \biggl(\frac{\overline{g}_{k}}{k\, \overline{f}_{k}}\biggr) \biggr], 
\nonumber\\
g_{k}(\tau) = A^{(r)}_{g\, f}(u,\, u_{r}) \, \overline{f}_{k} \, \biggl[ 1 + \frac{A^{(r)}_{g\,g}(u, \, u_{r}) }{A^{(r)}_{g\, f}(u,\, u_{r})} \biggl(\frac{\overline{g}_{k}}{k\, \overline{f}_{k}}\biggr) \biggr],
\label{SR0b}
\end{eqnarray}
where the second term appearing inside each of the squared brackets of Eq. (\ref{SR0b}) 
is always subleading in the limit $|u_{r}|\ll 1$. Thanks to Eq. (\ref{SR0b}) and bearing in mind the results 
of Eqs. (\ref{APPA1})--(\ref{APPA2}) the mode functions $F_{k}(u)$ and 
$G_{k}(u)$ are explicitly given by:
\begin{equation}
F_{k}(u) = \overline{F}_{k}^{(r)} j_{0}(u)\biggl[ 1 + {\mathcal O}(u_{r}^2)\biggr], \qquad \qquad G_{k}(u) = - k \, \overline{F}_{k}^{(r)} j_{1}(u)\biggl[ 1 + {\mathcal O}(u_{r}^2)\biggr],
\label{SR1}
\end{equation}
where $j_{0}(u)$ and $j_{1}(u)$ denote the (spherical) Bessel functions of order $0$ and $1$ respectively \cite{TRIC,ABR}. The amplitude of the mode functions $\overline{F}_{k}^{(r)}$ determines the amplitude of the tensor power spectrum during inflation for typical wavelengths larger than the Hubble radius, i.e. 
\begin{eqnarray}
\overline{P}_{T}^{(r)}&=& \frac{16}{\pi} \biggl(\frac{H_{r}}{M_{P}}\biggr)^2 \biggl(\frac{k}{a_{r} \, H_{r}}\biggr)^{n_{T}}
= \frac{128}{3} \biggl(\frac{V}{M_{P}^4}\biggr)_{k \simeq H_{r} a_{r}}, 
\label{SR2}
\end{eqnarray}
where we defined, for the sake of conciseness, $\overline{P}_{T}^{(r)}=\overline{P}_{T}(k,\tau_{r})$ and $n_{T} = - 2 \epsilon = - r_{T}/8$. In Eq. (\ref{SR2}) $V$ denotes the inflationary potential which is related to the expansion 
rate in the slow-roll approximation $3\, H^2 \overline{M}_{P}^2 \simeq V$; as already mentioned prior to Eq. (\ref{SR0b}) and in the related footnote $H_{r}$ is, roughly speaking, the expansion rate at the end of inflation. Note finally that $M_{P}$ (appearing in Eq. (\ref{SR2})), $\overline{M}_{P}$ and $\ell_{P}$ (introduced in Eq. (\ref{SR7})) are all related as $\overline{M}_{P} = \ell_{P}^{-1} = M_{P}/\sqrt{8 \, \pi}$.
All in all the autocorrelation functions of Eqs. (\ref{ST11})--(\ref{ST12}) and (\ref{ST13})--(\ref{ST14}) can be directly expressed in terms of $\overline{P}_{T}^{(r)}$
\begin{eqnarray}
\Delta_{k}(\tau_{1},\, \tau_{2}) &=& \frac{\pi^2 \, \overline{P}_{T}^{(r)}}{k^3} \, \frac{[\cos{(u_{1} - u_{2})} - \cos{(u_{1} + u_{2})}]}{u_{1}\, u_{2}},
\label{SR3}\\
\overline{\Delta}_{k}(\tau_{1},\, \tau_{2}) &=& \frac{\pi^2 \, \overline{P}_{T}^{(r)}}{k}\biggl[ \frac{\cos{(u_{1} - u_{2})}}{u_{1}\, u_{2}} \biggl( 1 + \frac{1}{u_{1}\, u_{2}}\biggr) + \frac{\sin{(u_{1} - u_{2})}}{u_{1} \, u_{2}} \biggl(\frac{1}{u_{2}} - \frac{1}{u_{1}}\biggr)
\nonumber\\
&+& \frac{\cos{(u_{1} + u_{2})}}{u_{1}\, u_{2}} \biggl( 1 - \frac{1}{u_{1}\, u_{2}}\biggr) -  \frac{\sin{(u_{1} + u_{2})}}{u_{1} \, u_{2}} \biggl(\frac{1}{u_{2}} + \frac{1}{u_{1}}\biggr)\biggr],
\label{SR4}
\end{eqnarray}
where, by definition, $u_{1} = u(\tau_{1})$ and $u_{2} = u(\tau_{2})$. But since at late-times $u_{1} \simeq k \tau_{1}$ and 
$u_{2} = k\tau_{2}$, the autocorrelation functions of Eqs. (\ref{ST9})--(\ref{ST10}) do not only depend on the time 
difference, as implied in the case of stationary processes discussed in section \ref{sec2}. On the contrary both autocorrelation 
functions $\Delta_{k}(\tau_{1},\, \tau_{2})$ and $\overline{\Delta}_{k}(\tau_{1},\, \tau_{2})$ include terms depending both on  $(\tau_{1} - \tau_{2})$ and on $(\tau_{1}  + \tau_{2})$. There also a number of corrections going as inverse powers of $u_{1}$
and $u_{2}$; some of these corrections are suppressed when the wavelengths of the gravitons are much smaller 
than the Hubble radius during the radiation stage.

\subsection{The autocorrelation functions during the matter stage}
The second relevant example involves the evolution during the matter stage. By enforcing at each stage of the evolution the  Wronskian normalization of Eq. (\ref{ST7e}) (and the related canonical form of the commutation relations) the values of the mode functions during the matter stage are given by
\begin{eqnarray}
f_{k}(\tau) = A^{(m)}_{f\, f}(v,\, v_{eq})\,  \overline{f}^{(m)}_{k} \, \biggl[1 + \frac{A^{(m)}_{f\,g}(v, \, v_{eq}) }{A^{(m)}_{f\, f}(v,\, v_{eq})} \biggl(\frac{\overline{g}^{(m)}_{k}}{k\, \overline{f}^{(m)}_{k}}\biggr) \biggr], 
\nonumber\\
g_{k}(\tau) = A^{(m)}_{g\, f}(v,\, v_{eq}) \, \overline{f}^{(m)}_{k} \, \biggl[ 1 + \frac{A^{(m)}_{g\,g}(v, \, v_{eq}) }{A^{(m)}_{g\, f}(v,\, v_{eq})} \biggl(\frac{\overline{g}^{(m)}_{k}}{k\, \overline{f}^{(m)}_{k}}\biggr) \biggr].
\label{SR7a}
\end{eqnarray}
In this case the elements of the transition matrix Eq. (\ref{SR7a}) have been listed in Eqs. (\ref{APPA3})--(\ref{APPA4}). Moreover the expression of $v= v(\tau)$ 
is the dust analog\footnote{The continuity of the scale factor and of its first conformal time derivative 
implies $a(\tau) = \{\beta(\epsilon) (\tau + \tau_{eq})+ 2 [\beta(\epsilon) +1]\tau_{r}\}^2/\{4 \tau_{r} [ \beta(\epsilon) \tau_{eq} + (\beta(\epsilon) +1)\tau_{r}]\}$ for $\tau \geq \tau_{eq}$.}
 of $u(\tau)$ introduced in Eq. (\ref{SR0})
\begin{equation}
v = v(\tau) = k [ \tau + \tau_{eq} + 2 ( 2 -\epsilon) \, \tau_{r}], \qquad \tau \gg \tau_{eq}.
\label{SR7}
\end{equation}
As in the case of $u_{r}$ we use here the shorthand notation $v_{eq} = v(\tau_{eq})$. By definition $\overline{f}_{k}^{(m)}$ and $\overline{g}_{k}^{(m)}$ are the values of the mode functions at the end of the radiation stage. From Eq. (\ref{SR0b})
we then have: 
\begin{eqnarray}
\overline{f}^{(m)}_{k} &=&A^{(r)}_{f\, f}(u_{eq},\, u_{r})\,  \overline{f}_{k} \, \biggl[ 1 + \frac{A^{(r)}_{f\,g}(u_{eq}, \, u_{r}) }{A^{(r)}_{f\, f}(u_{eq},\, u_{r})} \biggl(\frac{\overline{g}_{k}}{k\, \overline{f}_{k}}\biggr) \biggr], 
\nonumber\\
\overline{g}^{(m)}_{k} &=& A^{(r)}_{g\, f}(u_{eq},\, u_{r}) \, \overline{f}_{k} \, \biggl[ 1 + \frac{A^{(r)}_{g\,g}(u_{eq}, \, u_{r}) }{A^{(r)}_{g\, f}(u_{eq},\, u_{r})} \biggl(\frac{\overline{g}_{k}}{k\, \overline{f}_{k}}\biggr) \biggr],
\label{SR7aa}
\end{eqnarray}
where, recalling the expression of Eq. (\ref{SR0}),  $u_{eq} = u(\tau_{eq})\simeq k \tau_{eq}$.
As in the case of Eq. (\ref{SR0b}) the second term inside each of the squared brackets 
of Eq. (\ref{SR7a}) turns out to be subleading in the limit $|v_{eq}| \ll 1$ which is appropriate for all the wavelengths 
that are shorter than the Hubble radius during the dust stage. Using then the results 
of Eqs. (\ref{APPA3})--(\ref{APPA4}) together with Eqs. (\ref{SR7a})--(\ref{SR7aa}) the evolution of the mode functions during the matter stage takes the form
\begin{eqnarray}
F_{k}(v) &=& - \overline{F}_{k}^{(m)} \frac{j_{1}(v)}{v}\,\,\biggl[1 + {\mathcal O}(u_{r}^2) + {\mathcal O}(v_{eq}^2) + {\mathcal O}(u_{r}\, v_{eq})\biggr], 
\nonumber\\
G_{k}(v) &= & \overline{G}_{k}^{(m)} \frac{j_{2}(v)}{v} \,\,\biggl[1 + {\mathcal O}(u_{r}^2) + {\mathcal O}(v_{eq}^2) + {\mathcal O}(u_{r}\, v_{eq})\biggr],
\label{SR8}
\end{eqnarray}
where $j_{1}(v)$ and $j_{2}(v)$ are, respectively, the spherical 
Bessel functions of order $1$ and $2$. Finally the tensor power spectra 
$P_{T}(k,\tau)$ and $Q_{T}(k,\tau)$ can be directly expressed in terms 
of $\overline{P}_{T}(k,\tau_{r})$:
\begin{eqnarray}
&& P_{T}(k,\, v) = 9\,\overline{P}_{T}^{(r)}\,\biggl[ \frac{\cos{v}}{v^2} - \frac{\sin{v}}{v^3} \biggr]^2, 
\label{SR9}\\
\hspace{-1cm}
&& Q_{T}(k,\, v) =9\,k^2\,\overline{P}_{T}^{(r)} \,\biggl[ 3 \frac{\sin{v}}{v^{4}} - \frac{\sin{v}}{v^2} - 3 \frac{\cos{v}}{v^3} \biggr]^2,
\label{SR10}
\end{eqnarray}
implying that, in the limit $v \ll 1$, $P_{T}(k,\, v) \to \overline{P}_{T}(k,\tau_{r})$ and  $Q_{T}(k,\, v) \to k^2 \overline{P}_{T}(k,\tau_{r})$. The autocorrelation functions of Eqs. (\ref{ST13})--(\ref{ST14}) can therefore be expressed as
\begin{eqnarray}
\Delta_{k}(\tau_{1}, \, \tau_{2}) &=& \frac{9 \pi^2}{k^3} \frac{\overline{P}_{T}(k,\tau_{r})}{v_{1}^2 \, v_{2}^2}\biggl[ \cos{(v_{1} - v_{2})} \biggl( 1 + \frac{1}{v_{1}\, v_{2}}\biggr) + \sin{(v_{1} - v_{2})} \biggl(\frac{1}{v_{2}} - \frac{1}{v_{1}}\biggr)
\nonumber\\
&+&\cos{(v_{1} + v_{2})} \biggl( 1 - \frac{1}{v_{1}\, v_{2}}\biggr) -  \sin{(v_{1} + v_{2})} \biggl(\frac{1}{v_{2}} + \frac{1}{v_{1}}\biggr)\biggr],
\label{SR12}\\
\overline{\Delta}_{k}(\tau_{1}, \, \tau_{2}) &=& \frac{9 \pi^2}{k} \frac{\overline{P}_{T}(k,\tau_{r})}{v_{1}^2 \, v_{2}^2}\biggl[ \cos{(v_{1} - v_{2})} \, A_{-}(v_1, v_{2}) - \cos{(v_{1} + v_{2})} \, A_{+}(v_1, v_{2})
\nonumber\\
&+& 3 \sin{(v_{1} - v_{2})}\, B_{-}(v_1,\,v_2) + 3 \sin{(v_{1} + v_{2})}\, B_{+}(v_1,\,v_2)\biggr].
\label{SR13}
\end{eqnarray}
In Eqs. (\ref{SR12})--(\ref{SR13}) we use the notation $v_{1} = v(\tau_{1})$ and 
$v_{2} = v(\tau_{2})$. Furthermore the functions $A_{\pm}(v_{1},\, v_{2})$ and $B_{\pm}(v_{1},\, v_{2})$ are defined as:
\begin{eqnarray}
A_{\pm}(v_{1},\, v_{2}) = 1 - \frac{3}{v_{1}^2} - \frac{3}{v_{2}^2} + \frac{9}{v_{1}^2 \, v_{2}^2} \mp \frac{9}{v_1\, v_{2}}, \qquad
B_{\pm}(v_{1},\, v_{2}) = \frac{1}{v_{2}} \biggl(1 - \frac{3}{v_{1}^2}\biggr)
\pm \frac{1}{v_{1}}\biggl(1 - \frac{3}{v_{2}^2}\biggr).
\label{SR14}
\end{eqnarray}
As in the case of radiation, it is clear that the autocorrelation function 
does not simply depend on $(v_{1} -v_{2}) = k (\tau_{1} - \tau_{2})$
but also on $(v_{1} + v_{2}) = k[(\tau_{1} + \tau_{2}) + 2 \tau_{eq} + 4 ( 2 -\epsilon)]$. Moreover there are terms containing inverse powers of $v_{1}$ and $v_{2}$ which are not necessarily suppressed. The two examples proposed in the present and in the previous subsections 
can be extended to a various complementary situations 
where the intermediate expansion history deviates from the radiation 
dominated evolution and the high-frequency signal can be potentially much larger than in the case of the concordance paradigm \cite{EXP}.

\subsection{Asymptotic form of the autocorrelation functions}
When the wavelengths are all inside the Hubble radius the expressions of the autocorrelation
functions become simpler since all terms that are suppressed in the limit $u \gg 1$ and $v\gg 1$ 
can be neglected in the first approximation. In particular during the radiation dominated stage
for $u_{1} \simeq u_{2} \gg 1$ the contributions of the cosines always dominates in Eqs. (\ref{SR3})--(\ref{SR4}) so that 
\begin{eqnarray}
\Delta_{k}( \tau_{1},\, \tau_{2}) &=& \frac{\pi^2 \, \overline{P}_{T}^{(r)}}{k^3} \, \frac{[\cos{(u_{1} - u_{2})} - \cos{(u_{1} + u_{2})}]}{u_{1}\, u_{2}},
\label{SR5}\\
\overline{\Delta}_{k}(\tau_{1},\, \tau_{2}) &=& \frac{\pi^2 \, \overline{P}_{T}^{(r)}}{k} \, \frac{[\cos{(u_{1} - u_{2})} + \cos{(u_{1} + u_{2})}]}{u_{1}\, u_{2}},
\label{SR6}
\end{eqnarray}
where $(u_{1} - u_{2}) = k (\tau_{1} - \tau_{2})$ and $(u_{1} + u_{2}) = k [(\tau_{1} + \tau_{2}) + 2 ( 2 - \epsilon) \tau_{r}]$. 
During the dust-dominated epoch Eqs. (\ref{SR13})--(\ref{SR14}) can be evaluated in 
the limit $v_{1} \simeq v_{2} \gg 1$  and the result is formally similar to the 
one of Eqs. (\ref{SR5})--(\ref{SR6}) but with different phases:
\begin{eqnarray}
\Delta_{k}(\tau_{1}, \, \tau_{2}) &=& \frac{9 \pi^2}{k^3} \frac{\overline{P}_{T}^{(r)}}{v_{1}^2 \, v_{2}^2}\biggl[ \cos{(v_{1} - v_{2})}  + \cos{(v_{1} + v_{2})}\biggr],
\label{SR15}\\
\overline{\Delta}_{k}(\tau_{1}, \, \tau_{2}) &=& \frac{9 \pi^2}{k} \frac{\overline{P}_{T}^{(r)}}{v_{1}^2 \, v_{2}^2}\biggl[ \cos{(v_{1} - v_{2})}  - \cos{(v_{1} + v_{2})} \biggr].
\label{SR16}
\end{eqnarray}
The results of Eqs. (\ref{SR5})--(\ref{SR6}) and (\ref{SR15})--(\ref{SR16}) demonstrate
in explicit terms that a stationary stochastic process it is not equivalent, strictly 
speaking, to the random process that describes the relic gravitons at late times.
The reason for this lack of equivalence is related to the production mechanism: 
relic gravitons are produced in pairs of opposite three-momenta. This is why
the mode functions are ultimately standing waves; as a consequence the autocorrelation functions 
are not invariant under a common shift of the time coordinate. The possibility 
of observing the non-stationary nature of the autocorrelation function 
has been previously considered in the literature (see \cite{CORR5,CORR6} and 
references therein) with particular attention to the frequency widow of wide-band 
interferometers. If the frequency falls in the audio band we have that 
${\mathcal O}(100)\, \mathrm{Hz}$. Let us then consider the non-stationary contributions 
at the present time; the typical oscillation phase will be
\begin{equation}
v_{1} + v_{2} = k [(\tau_{1} + \tau_{2}) + 2 \tau_{eq}+ 4 (2 - \epsilon)\tau_{r}] = {\mathcal O}(\nu_{ref}/\nu_{p}),
\label{SR17}
\end{equation}
where $\nu_{ref}$ is a certain reference frequency and $\nu_{p} = {\mathcal O}(10^{-18})\, \mathrm{Hz}$. Even if $\nu_{ref}$ is sufficiently small the oscillation is so strong that the  contribution of $\cos{(v_{1} + v_{2})}$, possibly integrated over the instrumental window of sensitivity (e.g. between $\nu_{r}$ and a sufficiently close $\nu_{max}$), vanishes 
in comparison with the stationary contribution oscillating as $\cos{(v_{1} - v_{2})}$. In spite of this argument 
when the observations of the diffuse backgrounds exhibit non-stationary features it is probably wise to recall that 
the relic signal may also be non-stationary.
It is therefore essential to understand more accurately how the non-stationary corrections 
impact on the observables and, in particular, on the spectral energy density in critical units.

\renewcommand{\theequation}{5.\arabic{equation}}
\setcounter{equation}{0}
\section{The oscillations of the spectral energy}
\label{sec5}
The spectral amplitude $S_{h}(\nu)$ is not a well defined pivotal  variable for the description of the signal when the diffuse backgrounds are non-stationary. Following the discussion 
of section \ref{sec2} and taking into account the results of section \ref{sec4} it turns 
out that similar conclusions could hold also in the case of other widely employed quantities like the spectral energy density or the chirp amplitude. In spite of this generic expectation we are now going to argue that the spectral energy density, when correctly evaluated, is quasi-stationary since the strong oscillations (potentially present to leading order both in the power spectrum and in the chirp amplitude) are progressively more suppressed when the wavelengths of the gravitons become shorter than the Hubble radius at each corresponding epoch. For this purpose it is 
useful to remind that there are  actually two complementary ways in which the spectral energy density is usually evaluated, at least by looking at the current literature:
\begin{itemize}
\item{} in the first case the power spectrum is estimated (either analytically or numerically by means of an appropriate transfer function\footnote{The considerations 
reported here have a direct impact on the transfer functions that can be 
defined either for the tensor power spectrum \cite{TPS1,TPS2} or directly in terms of the spectral energy density \cite{relic2}. If the ultimate purpose is an accurate assessment 
of the spectral energy density, between the two strategies the latter is far more consistent than the former. Furthermore a direct numerical evaluation of the spectral 
energy density automatically smears the oscillations that are manually averaged when defining the transfer function with respect to the tensor power spectrum.}) and then related to the spectral energy density; this evaluation only requires the 
direct calculation of $P_{T}(k,\tau)$;
\item{} the second option is to estimate independently the power spectra $P_{T}(k,\tau)$ and $Q_{T}(k,\tau)$, compute the spectral energy density and then study the various relevant limits; among them the most relevant for the present ends is the one where all the wavelengths of the spectrum are shorter than the Hubble radius at each corresponding 
epoch.
\end{itemize} 
The first strategy turns out to be the limit of the second when the wavelengths are inside the Hubble radius but, as we shall see, the approximate expression where $\Omega_{gw}(k,\tau)$ 
only depends on $P_{T}(k,\tau)$  holds for the overall amplitude but not for the phases. This is why the first approach discussed above leads to an expression that is not stationary whereas the second one implies a result that is overall stationary (barring for the effects of the expansion of the background).  The quasi-stationarity comes, on a technical ground, from the destructive interference of the respective phases associated with $P_{T}(k,\tau)$ {\em and} $Q_{T}(k,\tau)$. The approximate and the exact approaches will now be analyzed in detail with the purpose of showing that the latter is more rigorous than the former. The spectral energy density turns then out to be  in practice, quasi-stationary and therefore particularly suitable for the description of the diffuse backgrounds of cosmic origin.

\subsection{Approximate evaluation and its spurious oscillations}
The heuristic strategy for the evaluation of the spectral energy density acknowledges that the spectral energy density is only determined by the tensor power spectrum $P_{T}(k,\tau)$ and that the phases of $P_{T}(k,\tau)$ coincide exactly with the ones of $\Omega_{gw}(k,\tau)$ according to 
 \begin{equation}
 \Omega_{gw}(k,\tau) = \frac{k^2}{12 {\mathcal H}^2}  \, P_{T}(k,\tau) \equiv \frac{k^2}{12\,a^2 \, H^2} P_{T}(k,\tau), \qquad 
 \biggl| \frac{k}{a \, H}\biggr| \gg 1.  
 \label{APPR1}
 \end{equation}
Equation (\ref{APPR1}) admittedly holds when the wavelengths are all shorter than the Hubble radius (i.e. $k \gg a\, H$ )but it also suggests that the strong oscillations of $P_{T}(k,\tau)$ are translated into the phases of $\Omega_{gw}(k,\tau)$. Let us now apply Eq. (\ref{APPR1}) to two illustrative examples. The first one involves the situation where the comoving wavelengths are all inside the Hubble radius during the radiation dominated stage. 
 \begin{figure}[!ht]
\centering
\includegraphics[height=7.5cm]{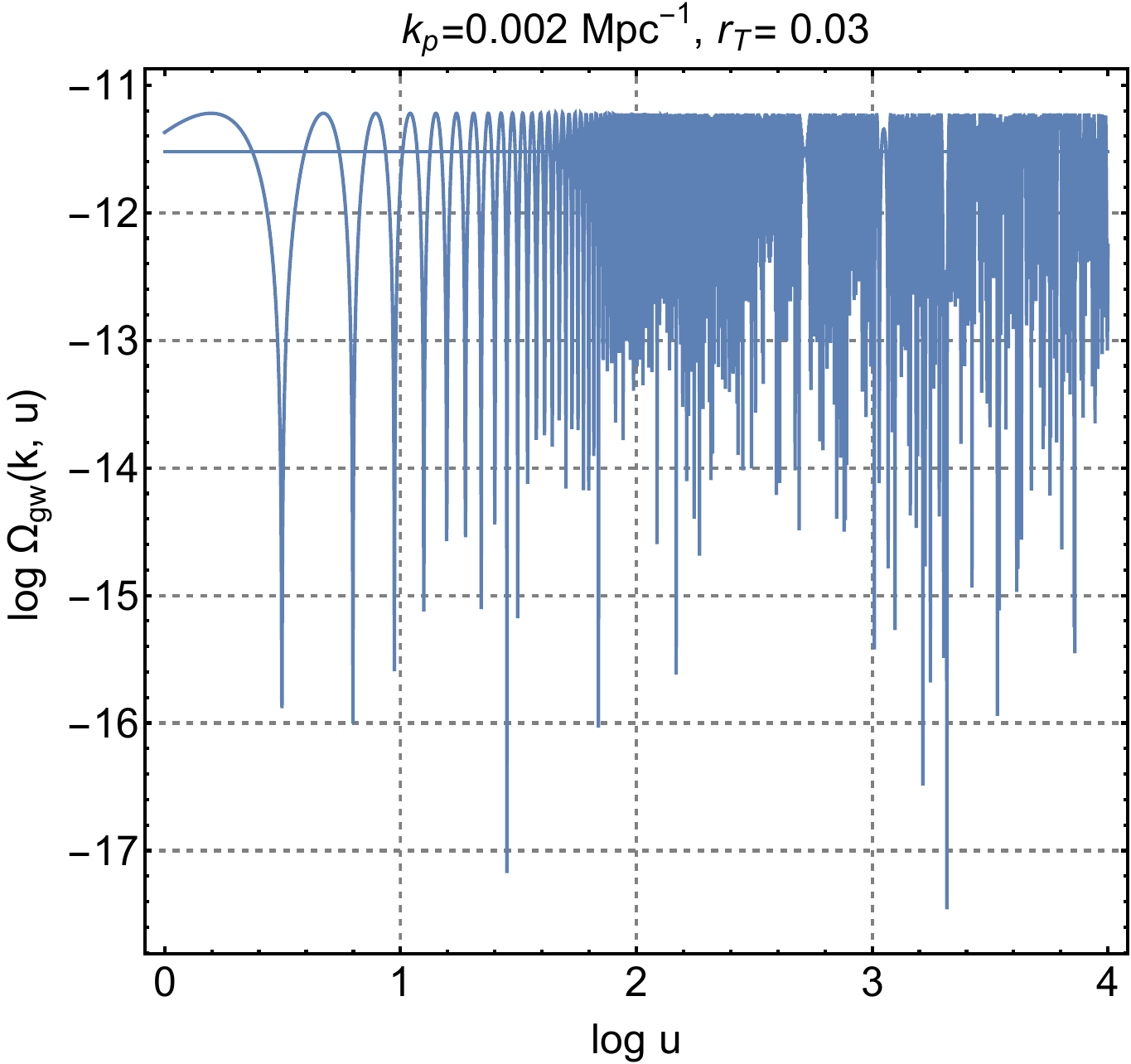}
\includegraphics[height=7.5cm]{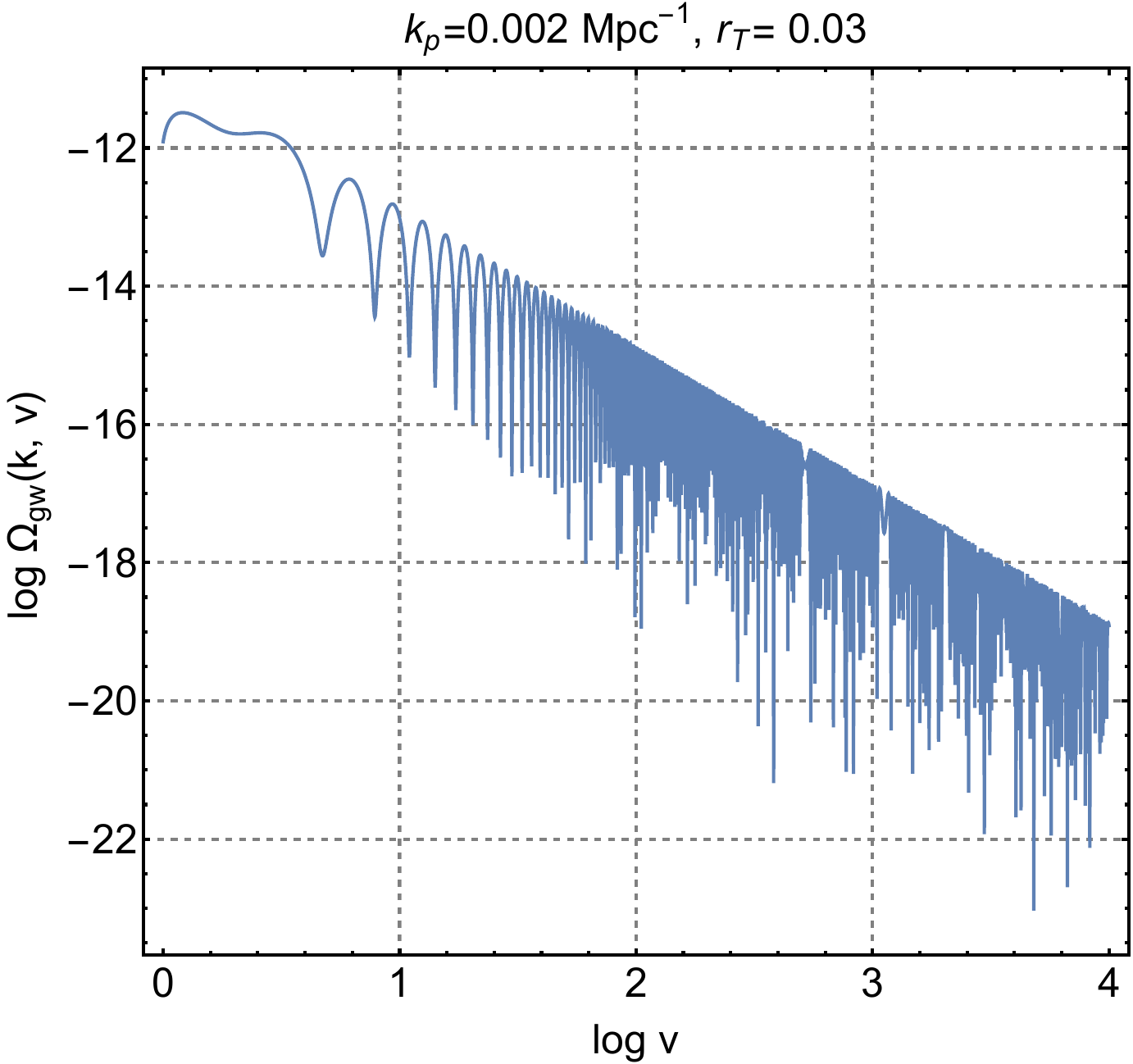}
\caption[a]{According to Eq. (\ref{APPR2}), the left plot we illustrate the common logarithm of $\Omega_{gw}(k,u)$ as a function of the common 
logarithm of $u$ (see Eq. (\ref{SR7})). In the plot at the right we illustrate instead the common logarithm of $\Omega_{gw}(k, v)$ as a function of the common logarithm of $v$ (see Eqs. (\ref{SR7}) and (\ref{APPR3})). Both results are a direct consequence 
of Eq. (\ref{APPR1}) which is, however, only approximate.}
\label{FIGURE1}      
\end{figure}
In this case inserting Eq. (\ref{SR0b}) into Eq. (\ref{APPR1}) we obtain\footnote{It is important to stress that the result of Eq. (\ref{APPR2}) refers to the spectral energy density computed during the radiation-dominated stage and not at the present time. Similar comments hold also for other quantities discussed in this section (see e. g. Eq. (\ref{APPR3})). } 
 \begin{eqnarray}
  \Omega_{gw}(k,u) &=&  \frac{u^2}{12} \overline{P}_{T}^{(r)}\, \, \biggl| 1 + \frac{A^{(r)}_{f\,g}(u, \, u_{r}) }{A^{(r)}_{f\, f}(u,\, u_{r})} \biggl(\frac{\overline{g}_{k}}{k\, \overline{f}_{k}}\biggr)\biggr|^2 
\nonumber\\
&=& \frac{\overline{P}_{T}^{(r)}}{12} \sin^2{u} \biggl[ 1 + {\mathcal O}(u_{r}^2)\biggr] \equiv \frac{\overline{P}_{T}^{(r)}}{12} \sin^2{u} \biggl[ 1 + {\mathcal O}(u_{r}^2)\biggr] \simeq \frac{\overline{P}^{(r)}_{T} }{24}  \,\, \bigl( 1 - \cos{2 u}\bigr), \qquad u \gg 1,
\label{APPR2}
\end{eqnarray}
where $\overline{P}^{(r)}_{T}$ has been already defined in Eq. (\ref{SR2}). The first 
equality in Eq. (\ref{APPR2}) follows from the exact result expressed in terms 
of the appropriate elements of the transition matrix (see also appendix \ref{APPB}). 
Overall the result of Eq. (\ref{APPR2}) 
 oscillates as $\sin^2{u}$ up to terms ${\mathcal O}(u_{r}^2)$: these are in practice the scales that left the Hubble radius during inflation and reentered in the radiation stage (i.e. for $k\tau_{r} < 1$). To be even more accurate the result of Eq. (\ref{APPR2}) applies for $u \gg 1$ and $\tau \gg \tau_{r}$ so that, eventually, the two conditions are also equivalent to $k \tau \gg 1$
and $k \gg a \, H$. The second expression of Eq. (\ref{APPR2}) clearly 
follows from the first one by applying standard trigonometric identities and to artificially 
get rid of the oscillating contributions some authors just average the obtained 
result over an oscillation period even if this is, strictly speaking, an arbitrary procedure. 

A further interesting example follows from the analysis of a dusty phase; if we actually insert the results of Eqs. (\ref{SR7a}) and (\ref{SR7aa}) into Eq. (\ref{APPR1}) we  obtain 
\begin{equation}
 \Omega_{gw}(k,v) = \frac{3}{16\, v^2} \overline{P}_{T}^{(r)} \biggl(\cos^2{v} + \frac{\sin^2{v}}{v^2} - \frac{\sin{2 v}}{v^3}\biggr) \biggl[ 1 + {\mathcal O}(u_{r}^2) + {\mathcal O}(v_{eq}^2) 
 + {\mathcal O}(u_{r}\, v_{eq})\biggr], \qquad v \gg 1.
 \label{APPR3}
 \end{equation}
As in the case of Eq. (\ref{APPR2}) the leading order result of Eq. (\ref{APPR3}) is strongly oscillating. In Fig. \ref{FIGURE1} we illustrate 
the results of Eqs. (\ref{APPR2})--(\ref{APPR3}). 
We recall that the amplitudes of the spectral energy density is controlled by 
$\overline{P}^{(r)}_{T}$ which can be parametrized as $ {\mathcal A}_{T} (k/k_{p})^{n_{T}}$
where $k_{p} = 0.002 \, \mathrm{Mpc}^{-1}$ is the pivot scale and ${\mathcal A}_{T} = r_{T} \, {\mathcal A}_{{\mathcal R}}$ the amplitude of the power spectrum. Since the tensor spectral index 
can be evaluated in terms of the consistency relations (i.e. $n_{T} = - r_{T}/8$) 
we can select, for instance, $r_{T} = 0.03$ (as suggested by the current data \cite{TS1,TS2,TS3}). 
The results of Fig. \ref{FIGURE1} refer to the case ${\mathcal A}_{\mathcal R} = 2.41 \times10^{-9}$ and for $k = {\mathcal O}(k_{p})$. While we purposely employed realistic figures for the various parameters, we stress nonetheless that the obtained expressions of the spectral energy density are not of direct phenomenological applicability. To be accurate the equality transition and a number of late-time effects must be numerically discussed \cite{relic}. The full form of the spectral energy density that shall be discussed in the following subsection 
is also potentially relevant for the numerical applications \cite{TS1,TS2,relic2}.
\begin{figure}[!ht]
\centering
\includegraphics[height=7.5cm]{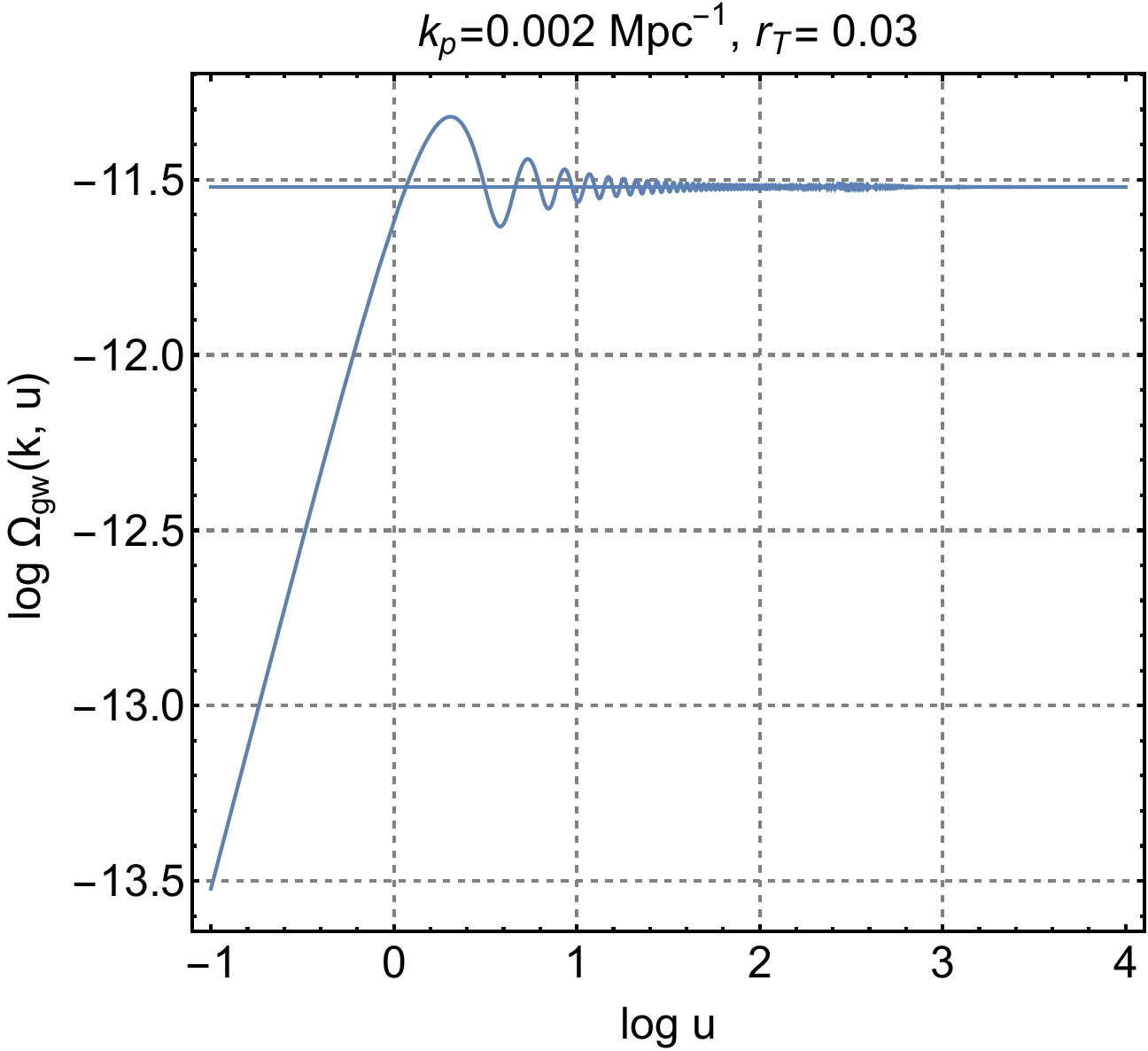}
\includegraphics[height=7.5cm]{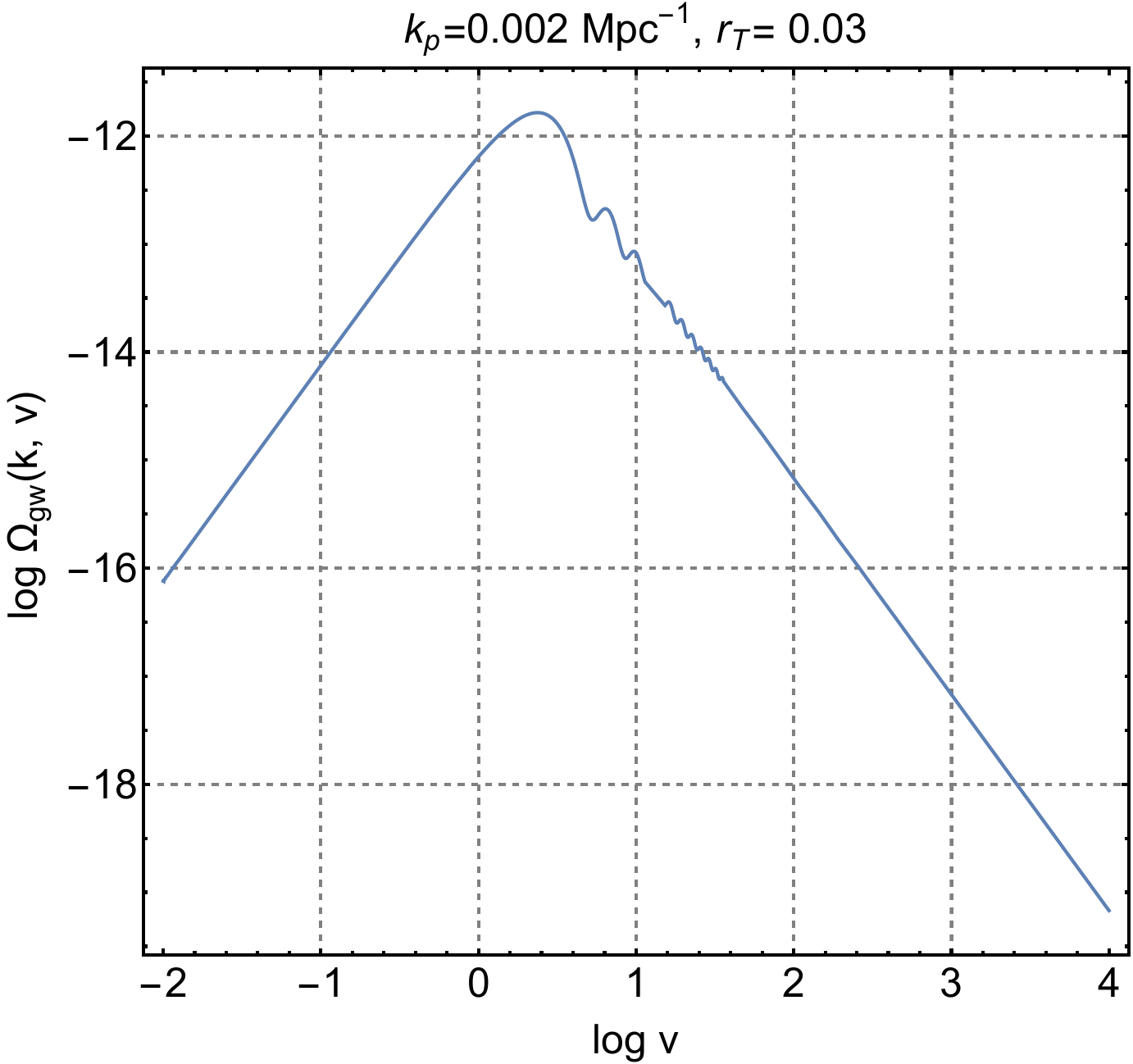}
\caption[a]{In the left plot we illustrate the common logarithm of $\Omega_{gw}(k, u)$ (given by Eq. (\ref{APPR7})) as a function of the common logarithm of $u$. In the right plot we 
report instead the common logarithm of the expression given 
in Eq. (\ref{APPR9}) as a function of the common logarithm of $v$. The results of the two 
plots should be compared with Fig. \ref{FIGURE1} which reports exactly the same quantities but computed from the approximate expression of Eq. (\ref{APPR1}). As already stressed in connection with Eq. (\ref{APPR2}) the results reported here do not apply at the present time and neglect a number of important damping sources like the neutrino free-streaming, the evolution of the relativistic species and the transition to the dominance of dark energy; see, in this respect, the discussion at the end of subsection \ref{subsec52}. }
\label{FIGURE2}      
\end{figure}

\subsection{Exact expression and the suppression of the oscillations}
\label{subsec52}
All in all we can summarize the results of the previous subsection by saying that Eq. (\ref{APPR1}) holds approximately only when the wavelengths are all much shorter than the Hubble radius. If the full form of the spectral energy density is instead adopted, the ${\mathcal O}(1)$ oscillations appearing in Fig. \ref{FIGURE1} practically disappear since they are suppressed for $k > a\, H$. Consequently their amplitude does not affect the leading term of the results and to clarify this point it is mandatory to start from a rigorous definition of the spectral energy density. For this purpose the energy density must be introduced from the variation of Eq. (\ref{ST15}) with respect to the background metric\footnote{Different definitions of the energy-momentum pseudo-tensor lead however to the same conclusion. The way the spectral energy density is assigned is actually not unique and this relevant point will be discussed at the end of this section. }; this procedure leads to a consistent energy-momentum pseudo-tensor of the relic gravitons \cite{GG3,GG4} (see also \cite{compar}):
\begin{equation}
T^{(gw)}_{\alpha\beta} = \frac{1}{4 \ell_{\mathrm{P}}^2} \biggl[ \partial_{\alpha} h_{i j} \,\,\partial_{\beta} h^{i j} 
- \frac{1}{2} \overline{g}_{\alpha \beta} \,\,\biggl(\overline{g}^{\mu\nu}\, \partial_{\mu} h_{ij} \,\,\partial_{\nu} h^{ij} \biggr)\biggr].
\label{APPR4}
\end{equation}
In Eq. (\ref{APPR4}) the indices of $T^{(gw)}_{\alpha\beta}$  are raised and lowered with the help of the background metric (i.e. $T_{\alpha}^{(gw)\,\, \beta} = \overline{g}^{\beta\nu} T^{(gw)}_{\alpha\nu}$). Equation (\ref{ST15}) implicitly assumes that the underlying background geometry is conformally flat so that in the case of a spatially flat Friedmann-Robertson-Walker metric $\overline{g}_{\mu\nu} = a^2(\tau)\, \eta_{\mu\nu}$
Eq. (\ref{APPR4}) leads directly to the energy-momentum tensor suggested by  Ford and Parker \cite{GG3,GG4} and the related energy density is \cite{compar}
\begin{equation}
\rho_{gw}(\vec{x}, \tau) = \frac{1}{8 \,\ell_{P}^2 \, a^2} \biggl( \partial_{\tau} h_{i\,j}\, \partial_{\tau}h^{i\,j} + \partial_{m} h_{i\,j} \partial^{m} h^{i\,j}\biggr).
\label{APPR5}
\end{equation}
If the field operators of Eqs. (\ref{ST4}) and (\ref{ST7}) are now inserted into Eq. (\ref{APPR5}) we can compute  the expectation value $ \langle \rho_{gw}(\vec{x}, \tau) \rangle$ so that the spectral energy density in critical units becomes:
\begin{equation}
\Omega_{gw}(k,\tau) = \frac{1}{\rho_{crit}} \frac{d \langle \rho_{gw} \rangle}{ d \ln{k}} = \frac{1}{24\, H^2 a^2} \biggl[ k^2 P_{T}(k,\tau) + Q_{T}(k,\tau)\biggr].
\label{APPR6}
\end{equation}
The evaluations of Eqs. (\ref{APPR2})--(\ref{APPR3}) can now be repeated
and compared with the results of Fig. \ref{FIGURE1}.  If we
consider a radiation-dominated stage and insert the results 
of Eqs. (\ref{SR0b})--(\ref{SR1}) inside Eq. (\ref{APPR6}) we obtain:
\begin{equation}
\Omega_{gw}(k, \,u) = \frac{1}{24}\,\,\overline{P}_{T}^{(r)}\,\,\biggl( 1 + \frac{\sin^2{u}}{u^2}  - \frac{\sin{2 u}}{u} \biggr).
\label{APPR7}
\end{equation}
The main difference between Eqs. (\ref{APPR2}) and (\ref{APPR7}) is that the former  oscillates
much more strongly than the latter. Furthermore while the expression of Eq. (\ref{APPR7}) 
applies both inside and outside the Hubble radius, Eq. (\ref{APPR2}) only applies when 
$u> 1$, i.e. for wavelengths shorter than the Hubble radius. For $\tau > \tau_{r}$ 
we have that, approximately, $u \simeq k \tau$ and when the wavelengths are shorter than the Hubble radius the spectral energy density is roughly constant up to 
oscillating corrections that are suppressed as $|k \tau|^{-1}$ and as $| k \,\tau|^{-2}$ in the limit 
$ |k\tau | \gg 1$, i.e.
\begin{equation}
\Omega_{gw}^{(r)}(k,\tau_{r}, \tau) =  \frac{\overline{P}_{T}(k, \tau_{r})}{24} \biggl( 1 +  {\mathcal O}\biggl(\frac{1}{k\tau}\biggr)+ 
{\mathcal O}\biggl(\frac{1}{k^2\tau^2}\biggr) \biggr), \qquad k \tau \gg 1.
\label{APPR8}
\end{equation}
From Eq. (\ref{APPR1}) the amplitude of oscillation is only determined from $P_{T}(k,\tau)$ 
but if we look at Eq. (\ref{APPR6}) we see that the  final results comes in fact from the 
mutual interference of $P_{T}(k,\tau)$ and $Q_{T}(k,\tau)$: while each of the terms 
is strongly oscillating their combination is quasi-stationary, as anticipated.
\begin{figure}[!ht]
\centering
\includegraphics[height=7.5cm]{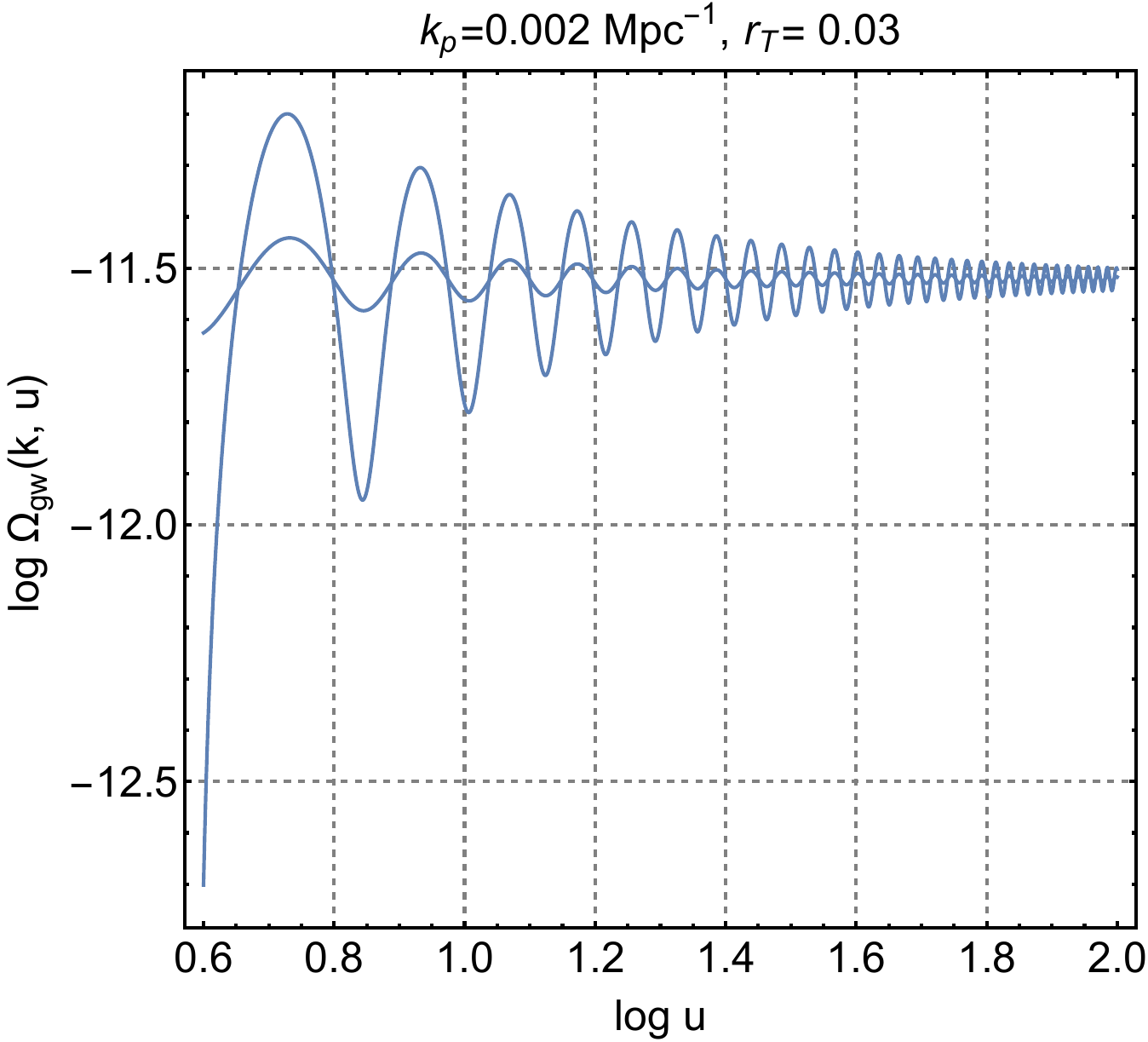}
\includegraphics[height=7.5cm]{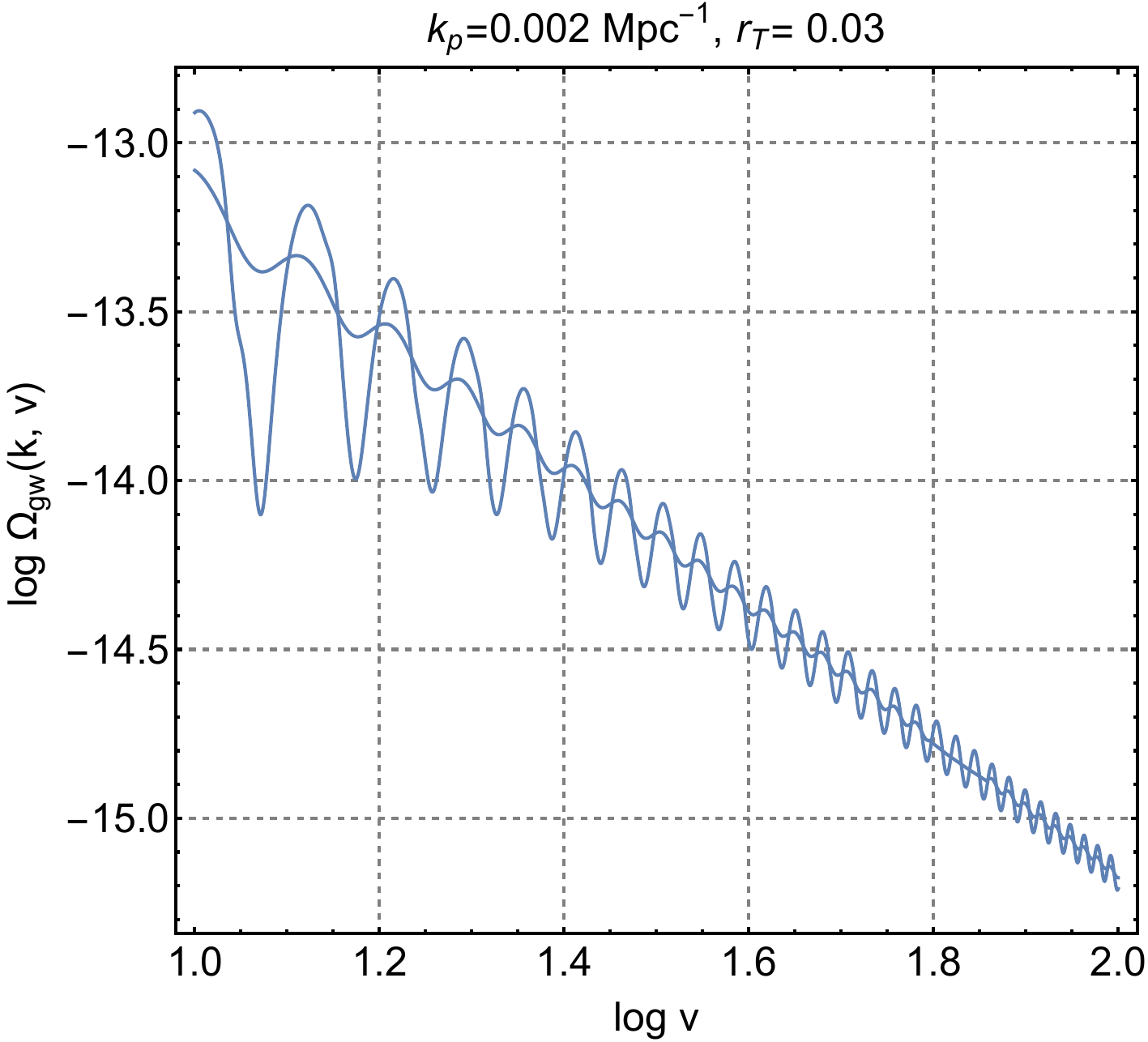}
\caption[a]{In the left plot the common logarithm of $\Omega_{gw}(k, u)$ (given by Eq. (\ref{APPR7})) is illustrated as a function of the common logarithm of $u$. In the right plot we 
report instead the common logarithm of the expression given 
in Eq. (\ref{APPR9}) as a function of the common logarithm of $v$. The left and the right plots should be compared with Fig. \ref{FIGURE1} that illustrates the approximate expression of Eq. (\ref{APPR1}). In both plots the oscillations with the largest amplitude follow from Eqs. (\ref{APPR7})--(\ref{APPR9}) while the ones with a comparatively smaller amplitude have been already given in Fig. \ref{FIGURE2}. We therefore conclude that different definitions 
of the energy-momentum pseudo-tensor modulate the final result in a different way but always 
suppress the oscillations that would appear to leading order.}
\label{FIGURE3}      
\end{figure}
The same approach adopted in the derivation of Eqs. (\ref{APPR7})--(\ref{APPR8}) can be applied 
during a dust-dominated stage with the result that the spectral energy density 
becomes
\begin{equation}
\Omega_{gw}(v) = \frac{3 \overline{P}_{T}^{(r)}}{32 \, v^2} \biggl[ 1 + 2 \frac{\sin{2 v}}{v} + 
\frac{ 9 \cos^2{v} - 5 \sin^2{v}}{v^2} - \frac{9 \sin{2 v}}{v^3} + 9 \frac{\sin^2{v}}{v^4} \biggr].
\label{APPR9}
\end{equation}
As in the case of Eq. (\ref{APPR8}), also the result of Eq. (\ref{APPR9}) holds both 
for $v < 1$ and $v >1$. Again, by comparing Eqs. (\ref{APPR3}) and (\ref{APPR9}) 
the strong oscillations of the leading term disappear and they get suppressed for $v\gg 1$. The results of Eqs. (\ref{APPR8})--(\ref{APPR9}) depend  on a pair of contributions coming, respectively, from $k^2 \, P_{T}(k,\tau)$  and $Q_{T}(k,\tau)$. 
The phases of oscillations appearing in Eq. (\ref{APPR8})  are controlled by the combination 
$[j_{0}^2(u) + j_{1}^2(u)]$ (where, as usual, $j_{0}(u)$ and $j_{1}(u)$ are the spherical Bessel functions).  When $u \gg 1$ we have that $j_{0}^2(u) \simeq j_{1}^2(u)$ 
so that we may replace $[j_{0}^2(u) + j_{1}^2(u)]$ with $2 j_{0}^2(u)$. This simplification 
correctly captures the amplitudes but {\em not} the phases of oscillation. 
The same argument applies in the case of Eq. (\ref{APPR9}) where the phases 
of oscillations are however controlled by the combination $[j_{1}^2(u) + j_{2}^2(u)]$. 
All in all the estimate of $\Omega_{gw}(k,\tau)$ inside the Hubble radius follows from 
Eq. (\ref{APPR6}) in the limit $k^2 P_{T}(k,\tau) \simeq Q_{T}(k,\tau)$ so that, at the very end, 
the spectral energy density in critical units becomes:
\begin{equation}
\Omega_{gw}(k,\tau) = \frac{k^2}{12\, H^2 a^2} \,\, P_{T}(k,\tau), \qquad k \gg a \, H,
\label{ENDD6}
\end{equation}
which coincides with Eq. (\ref{APPR1}).  Equation (\ref{ENDD6}) is grossly correct for typical  wavelength smaller than the Hubble radius (i.e. $k \gg a \, H$). But  Eq. (\ref{ENDD6}) {\em does not} correctly reproduce the  phases of the spectral energy density when the corresponding 
wavelengths are both shorter and larger than the Hubble radius. 
The results of Eqs. (\ref{APPR8}) and (\ref{APPR9}) are illustrated in Fig. \ref{FIGURE2}. The left plot of Fig. \ref{FIGURE2}
should be compared with the left plot of Fig. \ref{FIGURE1}: we clearly see that, in the limit $u > 1$ the oscillations  are suppressed in comparison with the evaluation based on Eq. (\ref{APPR1}). The horizontal line in both plots illustrates $\overline{P}_{T}^{(r)}/24$.

It is finally useful to stress that the results discussed in this section 
do not apply at the present time, as already mentioned in connection 
with Eqs. (\ref{APPR2})--(\ref{APPR3}). On the contrary the obtained 
results apply in the ranges defined by their respective arguments. As a general comment the spectral energy density at the present epoch is more suppressed by a factor that may range between $10^{-5}$ and $10^{-7}$ (see, for instance, \cite{relic}). This suppression is not only 
due to the redshift but also various damping sources including 
the neutron free streaming \cite{NNU1,NNU2,NNU3,NNU4}, the evolution of the relativistic species and the transition to the dominance of dark energy \cite{SS7}.  All these effect, however, are beyond the scope of the present discussion even if they may be crucial for a faithful 
assessment of the non-stationary features (especially over intermediate and high-frequencies) in concrete experimental situations.

\subsection{Complementary considerations}
 The different definitions of the 
energy density of gravitational waves \cite{onebb,twobb} stem directly from the equivalence principle that forbids the localization of the momentum of the gravitational field itself \cite{compar}. The same statement obviously holds for the pressure and the anisotropic stress of the gravitational waves. Since the energy-momentum tensor of the gravitational waves does not have a unique gauge-invariant (and frame-invariant) expression, through the years various expressions for the energy-momentum pseudo-tensor of the gravitational field have been deduced and they all lead to slightly different expressions of the energy density. Among them we may quote the Landau-Lifshitz 
approach \cite{LLexp1,LLexp2,LLexp3,LLexp4}, the Brill-Hartle strategy \cite{BHexp1}, the Isaacson pseudotensor \cite{BHexp2,BHexp3}. In this analysis we privileged the Ford-Parker energy-momentum tensor \cite{GG3} which is well defined both inside and outside the Hubble radius and it never leads to a violation of the energy conditions \cite{compar}. Other slightly different approaches have been pursued in the literature \cite{OTH1,OTH2,OTH3,OTH4,OTH5}. 

The purpose of this final subsection is to stress that the suppression
of the non-stationary features associated with the spectral energy density 
of the relic gravitons does not depend on the specific form of the pseudo-tensor.
The same kind of exercise discussed here can be repeated with all the 
various forms of the energy-momentum pseudo-tensor analyzed in Ref. \cite{compar}.
To give a swift example we then consider the explicit form of the Landau-Lifshitz 
pseudo-tensor which is defined from the second-order fluctuations of the Einstein tensor 
\cite{LLexp1,LLexp2,LLexp3} and its explicit form for a conformally flat metric $\overline{g}_{\mu\nu} = a^2(\tau)\eta_{\mu\nu}$ is given by:
\begin{equation}
\overline{\rho}_{gw}=  \frac{1}{8 a^2 \ell_{\rm P}^2} \biggl[ 8 {\mathcal H} \,\partial_{\tau}h_{i\,j }\,\, h^{i\,j} + \biggl( \partial_{m} h_{i\,j}\,\,\partial^{m} h^{i\,j} + 
\partial_{\tau} h_{i\,j}\,\, \partial_{\tau} h^{i\,j}\biggr)\biggr].
\label{APPR10}
\end{equation}
If we now compare the expression of $\rho_{gw}(\vec{x},\tau)$ 
given in Eq. (\ref{APPR5}) with the one of $\overline{\rho}_{gw}(\vec{x},\tau)$ 
we see that the difference is given by the first term inside the 
square bracket of Eq. (\ref{APPR10}). This term leads to a computable 
difference in the averaged energy density which is now given by
\begin{equation}
\langle \overline{\rho}_{gw}(\vec{x},\tau) \rangle =  \langle \rho_{gw}(\vec{x},\tau) \rangle+  \frac{2 {\mathcal H}}{ \, \pi^2 \, a^2 } \int k^2 \, d k  ( G_{k} \, F_{k}^{*} + G_{k}^{*} F_{k}).
\label{APPR11}
\end{equation}
As a consequence the spectral energy density in critical units will be different in the various expanding stages. In particular during a radiation-dominated stage of expansion we have that the spectral energy density 
$\overline{\Omega}_{gw}(k, u)$ is:
\begin{equation}
\overline{\Omega}_{gw}(k, u) = \frac{\overline{P}^{(r)}_{T}}{24}\biggl( 1 - \frac{7 \, \sin^2{u}}{u^2} + 3\frac{ \sin{2 u}}{u} \biggr).
\label{APPR12}
\end{equation}
If we now compare Eq. (\ref{APPR12}) with Eq. (\ref{APPR7}) we see that the last two terms 
inside the squared brackets have different coefficients. These terms, however, 
do not change the relevant conclusion of our discussion namely the absence 
of a dominant oscillating term. Furthermore it turns out that the oscillating 
contributions are also suppressed as the wavelengths get shorter than the Hubble 
radius. 

The comparison between the results of Eq. (\ref{APPR7}) and (\ref{APPR12}) 
is illustrated graphically in the left plot of Fig. \ref{FIGURE3}. We restricted 
the interval of $u$ to make the plot more readable; the curve with the largest 
amplitude of oscillation corresponds to the Landau-Lifshitz choice leading to Eq. (\ref{APPR12}).
The curve with a smaller amplitude of oscillation corresponds follows 
from Eq. (\ref{APPR7}). In both cases, as expected, the oscillating contributions are 
suppressed in the limit $u > 1$. The same analysis leading to the comparison of Eqs. (\ref{APPR7}) 
and (\ref{APPR12}) is now illustrated in the case of a dust-dominated evolution 
where $\overline{\Omega}_{gw}(k,u)$ can be written as: 
\begin{equation}
\overline{\Omega}_{gw}(v) = \frac{3 \overline{P}_{T}^{(r)}}{32 \, v^2} \biggl( 1 - 6 \frac{\sin{2 v}}{v} - \frac{14 + 25 \cos{ 2 v}}{v^2} + 39 \frac{\sin{2 v}}{v^3} 
- 39 \frac{\sin^2{v}}{v^4} \biggr).
\label{APPR13}
\end{equation}
The comparison of Eqs. (\ref{APPR9}) and (\ref{APPR13}) suggests once more 
that the coefficients of the terms appearing inside the squared bracket are different in the two cases even if the main physical conclusions are not modified since the oscillations are always suppressed in the limit $v > 1$. In the right plot of Fig. \ref{FIGURE3} the results of Eqs. (\ref{APPR9}) and (\ref{APPR13}) are graphically 
compared. The curve with the largest oscillatory amplitude corresponds to the 
Landau-Lifshitz pseudo-tensor. In both cases, however, the oscillations 
are suppressed in the limit $v > 1$ and this behaviour is crucially 
different from the one illustrated in Fig. \ref{FIGURE1}. The comparison shows that 
the oscillations appearing in Fig. \ref{FIGURE1} are only caused by the approximation
method and not by the dynamical evolution which is accurately described 
in terms of the spectral energy density.

\subsection{Some other variables and their quasi-stationary limits}
Having established that the spectral energy density is a quasi-stationary  
variable, it is natural to consider other variables whose oscillations 
may also be potentially large. If we consider, for instance, the tensor 
power spectrum and the chirp amplitude we see that they are strongly
oscillating functions when the wavelengths are shorter than the Hubble radius.
For concrete estimates the oscillations are arbitrarily averaged and often disregarded.
This heuristic approach can be improved by using the spectral energy density 
as pivotal variable: in this way the oscillations of all the other observables 
will be automatically smeared inside the Hubble radius. Consider, as an example, 
the radiation-dominated stage; in this case we have that the exact evaluation 
of $\Omega_{gw}(k, u)$ leads directly to Eq. (\ref{APPR7}). If we are now interested 
in the tensor power spectrum inside the Hubble radius (or in the chirp amplitude) 
we have  
\begin{eqnarray}
P_{T}(k, u) &=& \frac{\overline{P}_{T}^{(r)}}{2 \, u^2} \biggl(1 + 3 \frac{\sin{2 u}}{u} - 7 \frac{\sin^2{u}}{u^2}\biggr),
\label{OTH1}\\
h_{c}(k, u) &=& \frac{\sqrt{ \overline{P}_{T}^{(r)}}}{2 \, u} \sqrt{1 + 3 \frac{\sin{2 u}}{u} - 7 \frac{\sin^2{u}}{u^2}}.
\label{OTH2}
\end{eqnarray}
Equations (\ref{OTH1})--(\ref{OTH2}) are only valid when all the relevant wavelengths 
are inside the Hubble radius, i.e. $u > 1$ and $k \tau >1$. The same 
analysis can be repeated for  a dust-dominated stage and in all other 
physical situations. One of the points of the present analysis has been that the spectral amplitude 
cannot be used for an explicit evaluation of the signal. We maintain this point 
since the spectral amplitude is defined, strictly speaking, only in the case 
of a process that is truly stationary. However if we really want to use, for some 
reason, $S_{h}(\nu)$ also in a non-stationary situation we can use the same 
logic leading to Eqs. (\ref{OTH1})--(\ref{OTH2}): if we use the spectral energy density 
as pivotal variable we can always estimate 
\begin{equation}
\nu \, S_{h}(|\nu|) = \lim_{k\tau\gg 1} \frac{12 \, H^2 \, a^2}{k^2} \, \Omega_{gw}(k,\tau).
\label{OTH3}
\end{equation}
In the case of a radiation-dominated stage this strategy simply leads to 
\begin{equation}
\nu\, S_{h}(|\nu|) \to \frac{\overline{P}_{T}^{(r)}}{ 4 \pi^2 \nu^2 \tau^2}\biggl[ 1 + {\mathcal O}\biggl(\frac{a H}{ 2 \pi \nu}\biggr)\biggr].
\label{OTH4}
\end{equation}
In Eq. (\ref{OTH4}) the contribution of the expansion is always present but 
the oscillating contributions disappear from the leading term. Equation (\ref{OTH4}) 
does not imply that the relic gravitons lead to a stationary process but the 
oscillating contributions are suppressed for wavelengths much smaller than the Hubble radius. Since these wavelengths correspond to frequencies that are much larger than the present expansion rate it is not necessary anymore to average by hand strongly oscillating trigonometric contributions.  The considerations developed in this section are necessary for a sound construction of a template family for the relic graviton backgrounds. The potential signal must be accurately computed both in amplitude and slope since, as we showed, the shortcuts 
may enhance some of the spurious features that are simply related with the approximation methods. For the disambiguation between the relic and the late signals also the second-order correlation effects should be taken into due account \cite{relic3,relic4}.

\renewcommand{\theequation}{6.\arabic{equation}}
\setcounter{equation}{0}
\section{Concluding remarks}
\label{sec6}
When the autocorrelation functions of stationary and ergodic ensembles of random fields are evaluated at different times $\tau_{1}$ and $\tau_{2}$ they ultimately depend on the difference $|\tau_{1} - \tau_{2}|$. 
The gravitons produced quantum mechanically thanks to the early variation of the space-time curvature appear however from the inflationary vacuum with opposite comoving three-momenta. At the semiclassical level the quantum mechanical initial conditions represented by travelling waves turn into standing waves because of the evolution of the space-time curvature and this is why the diffuse backgrounds of relic gravitons are intrinsically non-stationary. The production of pairs with opposite momenta is then reflected into the standing oscillations that appear, with different features, in all the correlation functions and in the related observables. The lack of stationarity is reflected into the spectral energy density and in all the other observables that are customarily employed for the description of the relic signal. The first consequence of this observation is that the spectral amplitude cannot be used for a rigorous description of the signal since it is should be time-independent and determined, according to the Wiener-Khintchine theorem, by the Fourier transform of the autocorrelation function of the process. The non-stationary features of the diffuse backgrounds seem also reflected into the strong oscillations that characterize both the power spectra and the spectral energy density. 

The analysis of the present paper shows however that the spectral energy density is only mildly non-stationary since the time dependence associated with $\Omega_{gw}(\nu,\tau)$ turns out to be strongly suppressed in the large-scale limit provided a consistent definition of the energy density is adopted.  If, on the contrary, the spectral energy density is related to the tensor power spectrum within a more heuristic (but rather standard) perspective the oscillations dominate the leading terms. According to this second approximate strategy the spectral energy density and the tensor power spectrum are not related in general terms but only in the limit where all the comoving wavelengths are inside the Hubble radius at the present time. This investigation shows that the non-stationary nature of the process affects directly also the spectral energy density but in a much milder way which is furthermore suppressed in the large-scale limit.

If the spectral energy density is used as the pivotal variable the strong oscillations appearing 
in the power spectrum and in the chirp amplitude are smeared without 
assuming any ad hoc time average that is often employed in the description of the relic signal.
For a direct evaluation of the spectral energy density the optimal strategy is instead to take the large-scale limit {\em after} evaluating all the power spectra in their exact form. While the lack of stationarity of the relic graviton backgrounds is reflected into the time-dependence of the spectral energy density,  it is not true that the phases of oscillation of the tensor power spectrum are directly reflected in $\Omega_{gw}(\nu,\tau)$. The heuristic arguments suggesting that the phases of the spectral energy density and of the tensor power spectrum coincide has been clarified and partially refuted in this analysis. An accurate evaluation of the spectral energy density can be used to bridge the stationary and quasi-stationary descriptions of the relic gravitons. This means, in practice, that the relic signal cannot be simply identified by looking either at the slope or at the amplitude of a given observable. A  naive viewpoint stipulates that the detection of the relic gravitons should be achieved with instruments built for general purposes and by only looking at some features of the signal as if a physical template family for the diffuse backgrounds was just optional.  The ideas conveyed in this analysis show the opposite: while the spectral slopes of a stationary background may be confused with the relic gravitons, this  cannot happen if the slopes, the amplitudes and the correlation properties are concurrently analyzed in the construction of an appropriate non-stationary template. It is our opinion, as stressed in the past, that the  template family should wisely include also the second-order correlation effects that are a direct consequence of the quantum origin of the relic signal. 

\section*{Aknowledgements}
I wish to thank A. Gentil-Beccot, A. Khols, L. Pieper, S. Reyes, S. Rohr and J. Vigen of the CERN Scientific Information Service for their usual kindness during this investigation.

\newpage
\begin{appendix}

\renewcommand{\theequation}{A.\arabic{equation}}
\setcounter{equation}{0}
\section{Scalar random fields}
\label{APPA}
\subsection{Stationary processes}
In the case of scalar random fields the autocorrelation function is introduced 
in full analogy with the case of the random functions discussed in section \ref{sec2}.
In particular the autocorrelation function is be denoted as $\Gamma_{\phi}(\tau_{1} - \tau_{2})$ 
and it appears in the ensemble average of the scalar amplitudes:
\begin{equation}
\langle \phi(\widehat{k}, \, \tau_{1})\, \phi(\widehat{p}, \, \tau_{2}) \rangle = {\mathcal C}_{\phi}\, \delta^{(2)}(\widehat{k} - \widehat{p}) \,\, \Gamma_{\phi}(\tau_{1} - \tau_{2}).
\label{ST2e0}
\end{equation}
In Eq. (\ref{ST2e0}) the numerical constant ${\mathcal C}_{\phi}$ fixes the relation between power 
spectrum and the spectral amplitude; $\delta^{(2)}(\widehat{k} - \widehat{k}^{\prime}) = \delta(\varphi -\varphi^{\prime})\, \delta(\cos{\vartheta} - \cos{\vartheta}^{\prime})$ is the angular delta function. From Eq. (\ref{ST2e0}) the spectral amplitude is defined, in practice, as in the case of Eq. (\ref{ST2c})
\begin{equation}
\phi(\widehat{k},\nu) = \int_{-\infty}^{\infty} d\tau\,\, e^{- 2 i\, \pi \nu\tau} \, \phi(\widehat{k},\tau), \qquad \langle \phi(\widehat{k}, \nu)\,  \phi(\widehat{p}, \nu^{\prime}) \rangle = {\mathcal C}_{\phi} \,\delta(\nu + \nu^{\prime}) \, S_{\phi}(\nu) 
 \, \delta^{2}(\widehat{k} - \widehat{p}).
 \label{ST2f}
\end{equation}
The autocorrelation function $\Gamma_{\phi}(\tau_{1}- \tau_{2})$ and the spectral amplitude $S_{\phi}(\nu)$ are then related in a way similar to the one already established in Eq. (\ref{ST2d}):
\begin{equation}
S_{\phi}(\nu) = \int_{-\infty}^{+\infty} \, d z \, \Gamma_{\phi}(z) \, e^{ 2 i\, \pi\,\nu\, z}.
\label{ST2h}
\end{equation}
 From Eq. (\ref{ST2h}) we can also argue that to verify whether a given expression is the correlation function of a stationary random process we must find its Fourier transform and establish whether or not it is always positive semi-definite.  Equation (\ref{ST2h}) explicitly illustrates that if $\Gamma(|\tau_{1} - \tau_{2}|)$ is dimensionless $S_{\phi}(|\nu|)$ has dimensions of a time. In summary a scalar random field can therefore be represented in Fourier space as:
\begin{equation}
\phi(\vec{x}, \tau) = \int_{-\infty}^{\infty} d \nu \, \int d\,\widehat{k}  \, \, e^{ 2\,i\,\pi \, \nu\,( \tau - \widehat{k}\cdot\vec{x})}\, \,\phi(\nu, \widehat{k}), 
 \label{ST2e}
 \end{equation}
 where the angular integration is performed over $d\widehat{k} = d\cos{\vartheta} \, d\varphi$; if the field $\phi(\vec{x}, \tau)$ is real (as we shall assume throughout this discussion) then $\phi^{\ast}(\nu, \widehat{k}) = \phi(-\nu, \widehat{k})$. Equations (\ref{ST2f}) and (\ref{ST2e}) are fully consistent the starting point of Eq. (\ref{ST2e0}). 
 
 \subsection{Homogeneous processes}
An ensemble of scalar random fields described by Eqs. (\ref{ST2f})--(\ref{ST2e}) and 
characterized by a stationary autocorrelation function is also homogeneous. To consider this point in more detail we may actually compute the correlation function for separated spatial locations and 
\begin{equation}
\langle \phi(\vec{x}, \tau_{1}) \, \phi(\vec{y}, \tau_{2}) \rangle = 8 \pi {\mathcal C}_{\phi} \, \int_{0}^{\infty} S_{\phi}(|\nu|)\, e^{ 2 i\pi \nu\, (\tau_{1} - \tau_{2})}\, j_{0}( 2 \pi \, \nu\, r) \,\, d \nu, 
\label{ST2n}
\end{equation}
 and note that it always depends upon $r = | \vec{x} - \vec{y}|$. The result (\ref{ST2n}) follows from Eqs. (\ref{ST2f}) and (\ref{ST2e}); moreover,  since $j_{0}(z)$ denotes the  spherical Bessel function of zeroth order \cite{TRIC,ABR}, the result of Eq. (\ref{ST2n}) is well defined not only for $\tau_{1} \to \tau_{2}$ but also when $r\to 0$. A homogeneous random field can also be Fourier transformed in a slightly different manner, namely 
 \begin{equation}
 \phi(\vec{x}, \tau) = \frac{1}{(2\pi)^{3/2}} \int d^{3} k \,\, e^{-i \vec{k}\cdot \vec{x}} \, \overline{\phi}(\vec{k}, \tau), \qquad \qquad \overline{\phi}^{\ast}(\vec{k}, \tau) = \overline{\phi}(-\vec{k}, \tau).
 \label{ST2o}
 \end{equation}
If the stationarity of the process is disregarded, an ensemble of homogeneous scalar random fields must also obey
 \begin{equation}
 \langle \overline{\phi}(\vec{k}, \tau)\, \overline{\phi}(\vec{p}, \tau) \rangle = \frac{2 \pi^2}{k^3} \,\delta^{(3)}(\vec{k} + \vec{p}) \,\,  P_{\phi}(k,\tau),
 \label{ST2p}
 \end{equation}
 where $P_{\phi}(k,\tau)$ denotes, in the present notations, the scalar power spectrum; as anticipated 
 $P_{\phi}(k,\tau)$ (unlike the spectral amplitude) is always dimensionless. 
 Equations (\ref{ST2o})--(\ref{ST2p}) do not assume the stationarity 
 of the process but since Eqs. (\ref{ST2e}) and (\ref{ST2p}) are both valid Fourier representations of an ensemble of scalar random fields the two can be related. For this purpose we preliminarily compute from Eqs. (\ref{ST2o})--(\ref{ST2p}) the analog of Eq. (\ref{ST2n}) valid in the case $\tau_{1} \to \tau_{2} = \tau$:
 \begin{equation}
 \langle \phi(\vec{x}, \tau) \, \phi(\vec{y}, \tau) \rangle = \int_{0}^{\infty} \frac{d k}{k} \, P_{\phi}(k,\tau) \, j_{0}(k\,r),
  \label{ST2q}
 \end{equation}
 where, as before, $r = |\vec{x} - \vec{y}|$. Thus  Eqs. (\ref{ST2n}) and (\ref{ST2q}) must coincide in the two concurrent limits $\tau_{1} \to \tau_{2}$ and $r\to 0$
 \begin{equation}
 8 \pi {\mathcal C}_{\phi} \, \int_{0}^{\infty} S_{\phi}(|\nu|)\, d \nu = \int_{0}^{\infty} \frac{d k}{k} \, P_{\phi}(k,\tau). 
\label{ST2r}
\end{equation}
Equation (\ref{ST2r}) implies that by arranging the numerical factor ${\mathcal C}_{\phi}$ the spectral amplitude and the power 
spectrum coincide 
\begin{equation}
\nu \, S_{\phi}(|\nu|) = P_{\phi}(|\nu|), \qquad 8 \pi {\mathcal C}_{\phi} = 1.
\label{ST2s}
\end{equation}
provided $P_{\phi}(k,\tau)$ {\em does not have an explicit time dependence}. The tensor analog of Eq. (\ref{ST2s}) is given in Eq. (\ref{CONN1}) ad the numerical difference between the two conditions is only due to the sum over the tensor polarizations. According to Eq. (\ref{ST2s})  the connection between the power spectrum and the spectral amplitude is only well defined in the stationary case. On the other hand, if the process is only homogeneous (but not necessarily stationary) the power spectrum is always well defined but does not lead to an autocorrelation function 
that depends only on $(\tau_{1} - \tau_{2})$. 

\renewcommand{\theequation}{B.\arabic{equation}}
\setcounter{equation}{0}
\section{Transition matrices and their limits}
\label{APPB}
The elements of the transition matrix associated with a radiation dominated evolution are given by:
\begin{eqnarray}
&& A^{(r)}_{f\,f}(u, \, u_{r}) = \cos{(u - u_{r})} + \frac{\sin{(u - u_{r})}}{u_{r}}, 
\nonumber\\
&& A^{(r)}_{f\,g}(u, \, u_{r}) = \sin{(u - u_{r})},
\nonumber\\
&& A^{(r)}_{g\,f}(u, \, u_{r}) =\biggl(\frac{1}{u_{r}} - \frac{1}{u}\biggr) \cos{(u - u_{r})} - \biggl( 1 + \frac{1}{u\, u_{r}}\biggr) \sin{(u - u_{r})},
\nonumber\\
&&A^{(r)}_{g\,g}(u, \, u_{r}) = \cos{(u-u_{r})} - \frac{\sin{(u - u_{r})}}{u},
\label{APPA1}
\end{eqnarray}
where $u(\tau)$ has been introduced already in Eq. (\ref{SR0}) and, by definition, $u_{r} = u(-\tau_{r})$.
We furthermore note that $(u - u_{r}) = (\tau+ \tau_{r})$ and in the limit $\tau \to - \tau_{r}$ 
the off-diagonal terms of the transition matrix vanish while the diagonal ones 
tend to $1$. Deep in the radiation stage (i.e. $u \gg u_{r}$) the various entries 
of the transition matrix can also be expanded for $|u_{r}| \ll 1$ and, in this limit, 
Eq. (\ref{APPA1}) becomes:
\begin{eqnarray}
&& A^{(r)}_{f\,f}(u, \, u_{r}) = \biggl(\frac{u}{u_{r}}\biggr) \, j_{0}(u) + \frac{u\, u_{r}}{2} j_{0}(u) + 
{\mathcal O}(|u_{r}|^2),
\nonumber\\
&& A^{(r)}_{f\,g}(u, \, u_{r}) = \sin{u} - u_{r}  \cos{u}  + 
{\mathcal O}(|u_{r}|^2),
\nonumber\\
&& A^{(r)}_{g\, f}(u, \, u_{r}) = - \biggl(\frac{u}{u_{r}}\biggr) j_{1}(u) - \frac{u\, u_{r}}{2}  j_{1}(u) + {\mathcal O}(|u_{r}|^2),
\nonumber\\
&& A^{(r)}_{g\, g}(u, \, u_{r}) = - u\, j_{1}(u) - u\,  y_{1}(u) + {\mathcal O}(|u_{r}|^2),
\label{APPA2}
\end{eqnarray}
where $j_{n}(z)$ and $y_{n}(z)$ are the standard spherical Bessel functions 
of index $n$ and argument $z$. The spherical Hankel functions 
of first and second kind \cite{TRIC,ABR} are instead defined, respectively, as $h^{(1)}_{n}(z) = j_{n}(z) + i\, y_{n}(z)$ and as $h^{(2)}_{n}(z) = j_{n}(z) - i\, y_{n}(z)$. The transition matrix associated with the matter-dominated phase can be swiftly expressed in terms of the spherical Hankel functions:
\begin{eqnarray}
A^{(m)}_{f\,f}(v, \, v_{eq}) &=& \frac{i}{2} \, v\, v_{eq}\biggl[ h^{(1)}_{2}(v_{eq})\, h^{(2)}_{1}(v) - 
h^{(2)}_{2}(v_{eq}) h^{(1)}_{1}(v) \biggr],
\nonumber\\
A^{(m)}_{f\,g}(v, \, v_{eq}) &=& \frac{i}{2} \, v\, v_{eq}\biggl[ h^{(1)}_{1}(v_{eq})\, h^{(2)}_{1}(v) - 
h^{(2)}_{1}(v_{eq}) h^{(1)}_{1}(v) \biggr],
\nonumber\\
A^{(m)}_{g\,f}(v, \, v_{eq}) &=& \frac{i}{2} \, v\, v_{eq}\biggl[ h^{(2)}_{2}(v_{eq})\, h^{(1)}_{2}(v) - 
h^{(1)}_{2}(v_{eq}) h^{(2)}_{2}(v) \biggr],
\nonumber\\
A^{(m)}_{g\,g}(v, \, v_{eq}) &=& \frac{i}{2} \, v\, v_{eq}\biggl[ h^{(2)}_{1}(v_{eq})\, h^{(1)}_{2}(v) - 
h^{(1)}_{1}(v_{eq}) h^{(2)}_{2}(v) \biggr],
\label{APPA3}
\end{eqnarray}
where the dimensionless variable $v(\tau)$ has been defined in Eq. (\ref{SR7}); moreover, by definition, $v_{eq} = v(\tau_{eq})$.  As before the entries of the transition matrix given in Eq. (\ref{APPA3}) are all real. It can be immediately appreciated that in the limit $v \to v_{eq}$ the diagonal terms of the matrix go to $1$ while the two off-diagonal contributions vanish. During the matter-dominated epoch the results of Eq. (\ref{APPA3}) can be expanded in the limit $|v_{eq}| \ll 1$ and the explicit results become:
\begin{eqnarray}
A^{(m)}_{f\,f}(v, \, v_{eq}) &=& 3 \biggl(\frac{v}{v_{eq}^2}\biggr) \, j_{1}(v) + \frac{v}{2} j_{1}(v) + {\mathcal O}(|v_{eq}|^2),
\nonumber\\
A^{(m)}_{f\,g}(v, \, v_{eq}) &=& 3 \biggl(\frac{v}{v_{eq}}\biggr) \, j_{1}(v) + \frac{v \, v_{eq}}{2} j_{1}(v) + {\mathcal O}(|v_{eq}|^2),
\nonumber\\
A^{(m)}_{g\,f}(v, \, v_{eq})  &=& - 3 \biggl(\frac{v}{v_{eq}^2}\biggr)\, j_{2}(v) - \frac{v}{2} j_{2}(v) + {\mathcal O}(|v_{eq}|^2),
\nonumber\\
A^{(m)}_{g\,g}(v, \, v_{eq}) &=& -  \biggl(\frac{v}{v_{eq}}\biggr) \, j_{2}(v) - \frac{v\, v_{eq}}{2} j_{2}(v) + {\mathcal O}(|v_{eq}|^2).
\label{APPA4}
\end{eqnarray}
Both in the case of Eqs. (\ref{APPA1}) and (\ref{APPA4}) the commutation relations 
imply that the transition matrices must be unitary so that, in particular, 
$A^{(X)}_{f\,f}\, A^{(X)}_{g\,g} - A^{(X)}_{f\,g}\, A^{(X)}_{g\,f} =1 $ where 
$X= r,\, m$ and the various entries are all functions of their respective arguments 
in each of the corresponding stages; see also, in this respect, Eq. (\ref{SR0a2}) and the discussion
therein.

\end{appendix}

\end{document}